\documentclass[%
reprint,
superscriptaddress,
amsmath,amssymb,
aps,
pra,
floatfix,
]{revtex4-2}

\usepackage{tikz}

\usetikzlibrary{positioning}

\usetikzlibrary{quantikz2}
\usepackage{ifthen}
\usepackage{graphicx}%
\usepackage{dcolumn}%
\usepackage{bm}%
\usepackage{hyperref}%
\usepackage{booktabs}

\usepackage[caption=false]{subfig}
\usepackage{tabularx}
\usepackage{float}
\usepackage{xcolor}

\newcommand{\rev}[1]{\textcolor{red}{#1}}
\renewcommand{\rev}[1]{#1}
\definecolor{niceblue}{RGB}{55, 110, 215}
\definecolor{nicered}{RGB}{214, 39, 40}
\definecolor{nicegreen}{RGB}{98, 159, 116}
\definecolor{nicepurple}{RGB}{128, 0, 128} 

\usepackage[capitalize]{cleveref}

\begin{document}

\title{Scalable Simulation of Fermionic Encoding Performance on Noisy Quantum Computers}%

\author{Emiliia Dyrenkova}
\email{emiliia.dyrenkova@uwaterloo.ca}
\affiliation{Institute for Quantum Computing, University of Waterloo, Waterloo, ON, N2L 3G1, Canada}
\affiliation{David R. Cheriton School of Computer Science, University of Waterloo, Waterloo, ON, N2L 3G1, Canada}

\author{Raymond Laflamme}
\email{Deceased}
\affiliation{Institute for Quantum Computing, University of Waterloo, Waterloo, ON, N2L 3G1, Canada}
\affiliation{Perimeter Institute for Theoretical Physics, Waterloo, ON, N2L 2Y5, Canada}
\affiliation{Department of Physics and Astronomy, University of Waterloo, Waterloo, ON, N2L 3G1, Canada}

\author{Michael Vasmer}
\affiliation{Institute for Quantum Computing, University of Waterloo, Waterloo, ON, N2L 3G1, Canada}
\affiliation{Perimeter Institute for Theoretical Physics, Waterloo, ON, N2L 2Y5, Canada}

\begin{abstract}

A compelling application of quantum computers with thousands of qubits is quantum simulation.
Simulating fermionic systems is both a problem with clear real-world applications and a computationally challenging task.

In order to simulate a system of fermions on a quantum computer, one has to first map the fermionic Hamiltonian to a qubit Hamiltonian. 
The most popular such mapping is the Jordan-Wigner encoding, which suffers from inefficiencies caused by the high weight of some encoded operators. 
As a result, alternative \emph{local} encodings have been proposed that solve this problem at the expense of a constant factor increase in the number of qubits required.
Some such encodings possess local stabilizers, i.e., Pauli operators that act as the logical identity on the encoded fermionic modes.
A natural error mitigation approach in these cases is to measure the stabilizers and discard any run where a measurement returns a -1 outcome.
Using a high-performance stabilizer simulator, we classically simulate the performance of a local encoding known as the Derby-Klassen encoding and compare its performance with the Jordan-Wigner encoding and the ternary tree encoding.
Our simulations use more complex error models and significantly larger system sizes (up to $18\times18$) than in previous work.
We find that the high sampling requirements of postselection methods with the Derby-Klassen encoding pose a limitation to its applicability in near-term devices and call for more encoding-specific circuit optimizations.

\end{abstract}

\maketitle

\section{\label{sec:level1}Introduction}

It is widely believed that chemical and materials simulations will be a high-impact application of future quantum computers~\cite{cao2019quantumchemistry,bauer2020quantum,mcardle2020quantumcomputational,motta2022emergingquantum,dalzell2023quantumalgorithms}. 
For these applications the relevant degrees of freedom are often fermionic and therefore it is necessary to represent fermions using the building blocks of a quantum computer: qubits.
The first such representation (or encoding) was the Jordan-Wigner (JW) encoding~\cite{jordan1928ueber}, which has the advantage of efficiency in that it uses $n$ qubits to represent $n$ fermionic modes.
However, in the JW encoding local fermionic operators are often represented by non-local and high-weight qubit operators, which can require many gates to implement on hardware.

A variety of other fermionic encodings have been proposed in more recent times~\cite{bravyi2002fermionic,verstraete2005mapping,seeley2012bravyikitaev,havlicek2017operator,setia2019superfast,jiang2020optimal,derby2021compacta, algaba2025fermiontoqubitencodingsarbitrarycode, Algaba2024lowdepthsimulations}, some of which map local fermionic operators to local qubit operators at the cost of requiring more than one qubit per encoded fermion.
Some of these encodings have the property that the encoded states are contained in a subspace of the full Hilbert space, which allows us to detect errors by checking whether a state resides in this subspace.
And in certain encodings such as the Derby-Klassen (DK) encoding~\cite{derby2021compact}, the subspace is stabilized by local Pauli operators similar to those of topological codes~\cite{kitaev2003faulttolerant,bombin2006topological} and we can therefore detect errors by measuring these local operators.

Small-scale demonstrations of quantum chemistry have been carried out on quantum computers with promising results~\cite{omalley2016scalable,hempel2018quantum,googleaiquantumandcollaborators2020hartreefock,stanisic2022observing}.
But achieving quantum advantage in these tasks likely requires a larger number of qubits and deeper circuits than are possible with today's devices~\cite{clinton2024nearterm}.
Furthermore, predicting the performance of quantum computers in these tasks is fraught with difficulty due to our inability of simulating large and deep quantum circuits.
However, certain sub-theories of quantum mechanics, such as stabilizer quantum mechanics~\cite{gottesman1998heisenberg}, are classically simulable but even so exhibit quantum phenomena such as entanglement and superposition.
Indeed, stabilizer simulations are widely used to benchmark the performance of quantum error-correcting (QEC) codes under Pauli error models, and these simulations are believed to give a reliable indication of performance.

Here we use an efficient stabilizer circuit simulator~\cite{gidney2021stim} to benchmark the performance of fermionic encodings at sizes beyond all previous simulations.
We use the circuit error model, where every element (state preparation, gate, measurement) is followed by a stochastic Pauli error.
This error model is commonly used in QEC research as it models a large range of noise present in current devices.
Furthermore, arbitrary Markovian noise (including coherent errors) can be tailored into stochastic Pauli errors using randomized compiling~\cite{wallman2016noise,hashim2021randomized} with little or no experimental overhead~\cite{fruitwala2024hardware}.
Indeed, this technique has already proved effective in error mitigation techniques such as probabilistic error cancellation~\cite{vandenberg2023probabilistic,Ferracin2024efficiently}.

We leverage the error-detecting properties of the DK encoding to design robust procedures for preparing encoded Slater determinant states and measuring operators of the Fermi-Hubbard Hamiltonian.
We compare the performance of the DK encoding with the JW encoding and the ternary tree (TT) encoding~\cite{jiang2020optimal}, for circuits implementing first-order Trotterized time evolution in the 2D Fermi-Hubbard Hamiltonian. 
We also study the performance of multiple error mitigation methods for the DK encoding with random logical operators circuits inspired by~\cite{chien2023simulating}.
We find that, although promising in terms of potential accuracy improvements, the DK-enabled error mitigation suffers from high sampling costs that limit its applicability to very low physical error rates or small lattice sizes. 
Our results confirm the need for more specialized circuit optimizations akin to those implemented in~\cite{nigmatullin2024experimentaldemonstrationbreakevencompact}.

The remainder of this manuscript is structured as follows.
In \cref{sec:enc-f2q}, we give a brief overview of general fermionic encodings and review the definitions of the specific encodings that we study in this work.
In \cref{sec:circuits} we describe circuits for implementing simulations using certain fermionic encodings.
In \cref{sec:numerics} we present the results of our numerical study comparing the performance of the DK encoding with the JW encoding and the TT encoding.
And in \cref{sec:conclusion} we conclude by discussing possible extensions of our work including experiments on quantum hardware.

\section{Encoding fermions into qubits} \label{sec:enc-f2q}

\subsection{The encoding problem}

Fermionic systems are commonly described using second quantization operators, defined by
\begin{equation}
\{a_i, a_j^{\dagger}\} = a_i a_j^{\dagger} + a_j^{\dagger} a_i = \delta_{ij} \mathbb{I}, \quad \{a_i, a_j\} = \{a_i^{\dagger}, a_j^{\dagger}\} = 0.
\end{equation}
Indices $i \text{ and } j$ correspond to different fermionic modes. 
These operators capture the essential properties of fermionic systems through the following anticommutation relations:
\begin{equation}
(a_i)^2 = 0, \quad (a_i^{\dagger})^2 = 0, \quad a_i a_j = - a_j a_i.
\end{equation}
These properties restrict each mode to be either occupied or unoccupied (Pauli exclusion principle), and encode the antisymmetric nature of fermionic wavefunctions in Fock space.

In contrast, qubit systems use Pauli operators $X_i$, $Y_i$, and $Z_i$, which obey different commutation relations:
\begin{equation}
    [X_i,Y_i] = 2iZ_i,\quad [Y_i,Z_i] = 2iX_i,\quad [Z_i,X_i] = 2iY_i,
\end{equation}
and anticommutation relations:
\begin{equation}
    \{X_i,Y_i\} = \{Y_i,Z_i\} = \{Z_i,X_i\} = 0,\; X_i^2 = Y_i^2 = Z_i^2 = I_i.
\end{equation}
Here, the index $i$ corresponds to a qubit. Unlike fermionic operators, Pauli operators acting on different qubits always commute.

The most general way to define the fermionic encoding problem is as the problem of finding a set of Pauli operators that satisfy the anticommutation relations of creation and anihilation operators. 
Usually, one aims to make the locality, or weight, of the encoded operators as low as possible.
However, as will be demonstrated by the DK Encoding, designing an encoding for a particular Hamiltonian can optimize the weight beyond the optimal value of a general one-to-one encoding.

In this work we focus on the spinless Fermi-Hubbard model~\cite{hubbard1963electron,hubardhalfcentury}:
\begin{equation}
H = -t \sum_{\langle i,j \rangle} \left( a_i^{\dagger} a_j + a_j^{\dagger} a_i \right) + U \sum_{\langle i,j \rangle} a_i^\dagger a_i a_j^\dagger a_j,
\end{equation}
where the set of neighbors $\langle i,j \rangle$ defines the interaction graph of the fermionic system, which we take to be a 2D square lattice.
The term $a_i^{\dagger} a_j + a_j^{\dagger} a_i$ is traditionally called the hopping term, while $a_i^\dagger a_i a_j^\dagger a_j$ is referred to as the Coulomb interaction term.

\subsection{Jordan-Wigner encoding}

The most famous and conceptually simple fermionic encoding is the one based on Jordan-Wigner transformation~\cite{jordan1928ueber}, which is defined as follows:
\begin{align}
a_j &= \left( \prod_{k=1}^{j-1} Z_k \right) \frac{X_j - i Y_j}{2}\\
a_j^{\dagger} &= \left( \prod_{k=1}^{j-1} Z_k \right) \frac{X_j + i Y_j}{2}
\end{align}
where \( X_j \), \( Y_j \), and \( Z_j \) are the Pauli operators acting on the \( j \)-th qubit. 

Without loss of generality, assume that $i<j$.
Under this encoding the hopping term of the Fermi-Hubbard Hamiltonian becomes
\begin{equation}
a_i^{\dagger} a_j + a_j^{\dagger} a_i \rightarrow \frac{1}{2} \left( \prod_{k=i+1}^{j-1} Z_k \right) \left( X_i X_j + Y_i Y_j \right),
\end{equation}
and the  Coulomb interaction term of Fermi-Hubbard Hamiltonian becomes
\begin{equation} \label{eq:jw-coulomb}
a_i^\dagger a_i a_j^\dagger a_j \rightarrow \frac{1}{4} (I + Z_i Z_j - Z_i - Z_j).
\end{equation}

\subsection{Ternary Tree encodings}

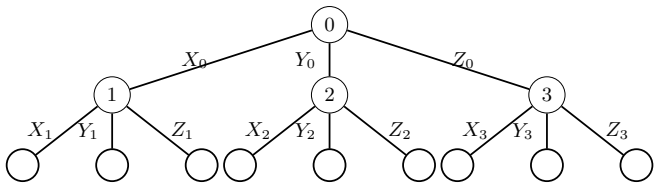
\begin{figure}[ht]
    \centering
     \scalebox{0.85}{
\begin{tikzpicture}[
    level 1/.style={sibling distance=34mm},
    level 2/.style={sibling distance=14mm},
    every node/.style={draw, circle, minimum size=5mm, text centered, font=\small},
    edge from parent/.style={draw, thick},
    level distance=11mm
]

\node {0} %
    child {node {1} %
        child {node {} edge from parent node[left, draw=none] {\(X_1\)}} %
        child {node {} edge from parent node[left, draw=none] {\(Y_1\)}} %
        child {node {} edge from parent node[right, draw=none] {\(Z_1\)}} %
        edge from parent node[left, draw=none] {\(X_0\)}} %
    child {node {2} %
        child {node {} edge from parent node[left, draw=none] {\(X_2\)}} %
        child {node {} edge from parent node[left, draw=none] {\(Y_2\)}} %
        child {node {} edge from parent node[right, draw=none] {\(Z_2\)}} %
        edge from parent node[left, draw=none] {\(Y_0\)}} %
    child {node {3} %
        child {node {} edge from parent node[left, draw=none] {\(X_3\)}} %
        child {node {} edge from parent node[left, draw=none] {\(Y_3\)}} %
        child {node {} edge from parent node[right, draw=none] {\(Z_3\)}} %
        edge from parent node[right, draw=none] {\(Z_0\)}}; %

\end{tikzpicture}
}
\caption{Ternary Tree Encoding operators are defined by assigning qubit indices to tree nodes that are not leaves and Pauli operators to the edges. This example tree creates 9 anticommuting Pauli operators on 4 qubits, with any subset of 8 of them giving a set of 8 encoded Majorana operators for 4 fermionic modes.}
\label{fig:ter-tree}
\end{figure}

At this point, it is useful to define the Majorana operators for a given fermionic mode \( j \):
\begin{equation}
\gamma_{2j} = a_j + a_j^{\dagger}, \quad \gamma_{2j+1} = -i (a_j - a_j^{\dagger}).
\end{equation}
Majorana operators are an equivalent representation of fermionic systems that satisfy simpler anticommutation relations:
\begin{equation}
\{\gamma_u, \gamma_v\} = 2\delta_{uv}, \quad \text{for } u, v \in [2n],
\label{eq-maj}
\end{equation}
where we use the shorthand $[2n]$ to denote $\{1, 2, \ldots, 2n \}$.

Any set of $2n$ Pauli operators that satisfy \cref{eq-maj} defines a valid encoding. 
Ternary trees have been used to find optimal encodings in terms of the weight of encoded Majorana operators. 
To demonstrate this, consider a small example in \cref{fig:ter-tree}. 
Each numbered node corresponds to a qubit with $n_{\text{modes}}=n_{\text{qubits}}=4$. 
Now, all paths from root to leaf correspond to mutually anticommuting Pauli strings by construction.
Since we need 8 encoded Majorana operators for $n_{\text{modes}}=4$, we can select any 8 Pauli operators from the tree.
In~\cite{jiang2020optimal} it was proved that the average weight $w$ of a one-to-one encoding must satisfy $w\geq \log_3{2n}$, with ternary tree encodings saturating the bound. 
In our simulations, we use the pruned Sierpinski tree encoding from~\cite{harrison2024sierpinski}. 

The ternary tree is defined differently for different sizes of the Hamiltonian, therefore the encoding of specific Fermi-Hubbard operators will depend on the size of the system represented.

\subsection{Derby-Klassen encoding}

The Derby-Klassen (DK) or Compact Encoding~\cite{derby2021compacta} is an encoding optimized for the 2D square lattice spinless Fermi-Hubbard model. 

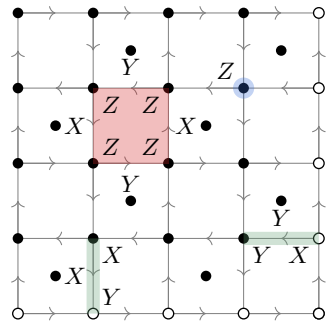
\begin{figure}[ht]
    \centering
    \begin{tikzpicture}
      \draw[step=1cm,gray,very thin] (0,0) grid (4,4);
      \foreach \x in {0,1,2,3} {
        \foreach \y in {0,1,2,3,4} {
          \pgfmathparse{int(mod(\y, 2))}
          \ifnum\pgfmathresult=0
          \draw[->,gray] (\x,\y) -- (\x+0.5,\y);
          \else
          \draw[->,gray] (\x+1,\y) -- (\x+0.5,\y);
          \fi
        }
      }
      \foreach \x in {0,1,2,3,4} {
        \foreach \y in {0,1,2,3} {
          \pgfmathparse{int(mod(\x, 2))}
          \ifnum\pgfmathresult=0
          \draw[->,gray] (\x,\y) -- (\x,\y+0.5);
          \else
          \draw[->,gray] (\x,\y+1) -- (\x,\y+0.5);
          \fi
        }
      }
      \foreach \x in {0,1,2,3} {
        \foreach \y in {1,2,3,4} {
          \fill (\x,\y) circle (2pt);
        }
      }
    \foreach \y in {0,1,2,3,4} {
        \draw[fill=white,line width=0.5pt] (4,\y) circle (2pt);
    }
    \foreach \x in {0,1,2,3,4} {
        \draw[fill=white,line width=0.5pt] (\x,0) circle (2pt);
    }
      \foreach \x in {0,1,2,3} {
        \foreach \y in {0,1,2,3} {
          \pgfmathparse{int(mod(\x + \y, 2))}
          \ifnum\pgfmathresult=0
          \fill (\x+0.5,\y+0.5) circle (2pt);
          \fi
        }
      }
      \fill[nicered,opacity=0.3] (1,2) rectangle (2,3);
      \node[anchor=south west] at (1,2) {$Z$};
      \node[anchor=south east] at (2,2) {$Z$};
      \node[anchor=north west] at (1,3) {$Z$};
      \node[anchor=north east] at (2,3) {$Z$};
      \node[anchor=south] at (1.5,1.5) {$Y$};
      \node[anchor=north] at (1.5,3.5) {$Y$};
      \node[anchor=west] at (0.5,2.5) {$X$};
      \node[anchor=east] at (2.5,2.5) {$X$};
      \fill[niceblue,opacity=0.3] (3,3) circle (4pt);
      \node[anchor=south east] at (3,3) {$Z$};
      \draw[line width=5pt,nicegreen,opacity=0.3] (3,1) -- (4,1);
      \node[anchor=north] at (3.5,1.5) {$Y$};
      \node[anchor=north east] at (4,1) {$X$};
      \node[anchor=north west] at (3,1) {$Y$};
      \draw[line width=5pt,nicegreen,opacity=0.3] (1,0) -- (1,1);
      \node[anchor=north west] at (1,1) {$X$};
      \node[anchor=south west] at (1,0) {$Y$};
      \node[anchor=west] at (0.5,0.5) {$X$};
    \end{tikzpicture}
    \caption{DK encoding lattice of size $4 \times 4$ with periodic boundaries. Each hollow vertex is identified with the corresponding filled vertex on the opposite boundary. 
    We highlight a stabilizer in red, a vertex operator in blue, and two edge operators in green.}
    \label{fig:ce-lattice}
\end{figure}

Let $\mathcal L$ be a square lattice of length $L \times L$ with periodic boundary conditions.
We use $\mathcal L_v$, $\mathcal L_e$, and $\mathcal L_f$ to refer to the vertices, edges, and faces of $\mathcal L$, respectively.
Recall that the faces of the square lattice are two-colourable, i.e., each face can be assigned a colour such that faces sharing an edge have different colours.
We call these two sets of faces red and blue, and we denote them by $\mathcal L_f^r$ and $\mathcal L_f^b$, respectively.
In the DK encoding, we place a qubit on each vertex and each blue face of $\mathcal L$.
Each edge is also assigned an orientation as shown in \cref{fig:ce-lattice}.

The DK encoding has an encoded fermionic mode for every vertex of $\mathcal L_v$ and is customarily defined in terms of a convenient Majorana operator basis:
\begin{equation}
\begin{split}
     V_j &= -i \gamma_{2j} \gamma_{2j+1} \quad j \in \mathcal L_v, \\
     E_{ij} &= -i \gamma_{2i} \gamma_{2j} \quad (i,j) \in \mathcal L_e,
\end{split}
\end{equation}
where $\gamma_{2j}$ and $\gamma_{2j+1}$ are Majorana operators at vertex $j$, and we refer to $V_j$ and $E_{ij}$ as vertex and edge operators, respectively.
The DK encoding maps these operators to the following Pauli operators:
\begin{equation}
    V_j = Z_j,
\end{equation}
\begin{equation}
\begin{split}
&E_{ij} = 
\begin{cases} 
X_i Y_j X_{f(i, j)} & \quad (i, j) \text{ oriented downwards}, \\
-X_i Y_j X_{f(i, j)} & \quad (i, j) \text{ oriented upwards}, \\
X_i Y_j Y_{f(i, j)} & \quad (i, j) \text{ horizontal},
\end{cases} \\
&E_{ji} = -E_{ij}
\end{split}
\label{eq:edge-ops-ce}
\end{equation}
where indices $i,j$ correspond to vertex qubits, and $P_{f(i,j)}$ denotes a Pauli operator $P$ acting on the qubit located at the face containing edge $(i,j)$. 
In our simulations, we only consider encodings defined on lattices with even $L$, as this ensures that there are no undetectable Majorana errors~\cite{derby2021compact}.

This encoding maps the hopping terms of the Fermi-Hubbard Hamlitonian as follows:
\begin{align}
a_i^{\dagger} a_j + a_j^{\dagger} a_i \to 
\begin{cases} 
\frac{1}{2} \left( X_{i} X_{j} Y_{f_{(i,j)}} + Y_{i} Y_{j} Y_{f_{(i,j)}} \right), \\ (i,j)\text{ horizontal}, \\[10pt]
\frac{1}{2} (-1)^{g(i,j)} \left( X_{i} X_{j} X_{f_{(i,j)}} + Y_{i} Y_{j} X_{f_{(i,j)}} \right), \\ (i,j)\text{ vertical},
\end{cases}
\label{eq:FH-CE-hop}
\end{align}
where $g(i,j)$ allows for a factor of $(-1)$ when appropriate due to the definition of vertical edge operators in \cref{eq:edge-ops-ce}. 
The encoding for the Coulomb terms is the same as \cref{eq:jw-coulomb}.

The DK encoding has stabilizers corresponding to closed loops of Majorana operators~\cite{derby2021compacta,jiang2019majorana}.
We consider a generating set of the stabilizers comprising operators associated with the red faces of $\mathcal L$.
Let $k \in [L^2/2]$ index the red faces.
For each $k$ we have a stabilizer generator $S_k$ formed by taking the product of all edge operators surrounding face $k$. Each of these operators is weight eight and acts on the vertex qubits surrounding the face as $Z$, on horizontally neighbouring face qubits as $X$, and on vertically neighbouring face qubits as $Y$.

The explicit form of such a stabilizer generator with the vertex qubits surrounding face $k$ indexed as $i,j,l,m$ in clockwise direction starting from upper left corner:
\begin{equation}
    S_{k} = Z_iZ_jZ_lZ_mY_{f_{(i,j)}}Y_{f_{(l,m)}}X_{f_{(j,l)}}X_{f_{(m,i)}}
\end{equation}
See \cref{fig:ce-lattice} for an example.

The code distance of the DK is equal to one, but despite this the stabilizers detect single-qubit $X$ and $Y$ errors as well as some higher weight errors~\cite{bausch2020mitigating}.
And in certain contexts the undetectable single-qubit $Z$ errors act as natural noise in the simulated system~\cite{bausch2020mitigating}. 

\section{Circuits for fermionic encodings} \label{sec:circuits}

In this section, we describe circuits for performing simulations using the JW, DK, and ternary encodings.
For the DK encoding, the stabilizer structure allows us to construct circuits that are robust to any single fault during their execution.

\subsection{Stabilizer measurement}
\label{subsec:stab-meas}

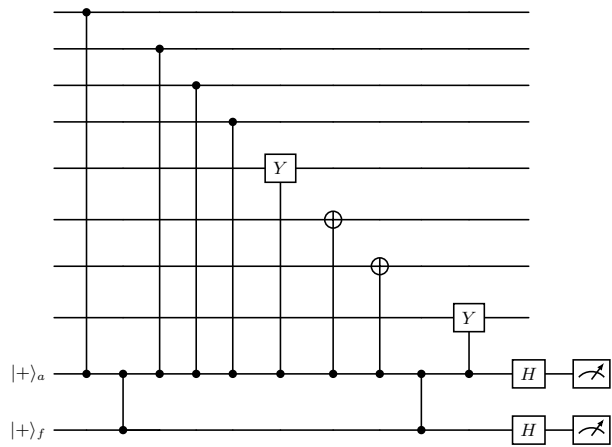
\begin{figure}[ht]
    \centering
    \begin{tikzpicture}
    \node[scale=.75] {
    \begin{quantikz}
    & \control{} & & & & & & & & & & \\
    & & & \control{} & & & & & & & & \\
    & & & & \control{} & & & & & & & \\
    & & & & & \control{} & & & & & & \\
    & & & & & & \gate{Y} & & & & & \\
    & & & & & & & \targ & & & & & \\
    & & & & & & & & \targ & & & & \\
    & & & & & & & & & & \gate{Y} & \\
    \lstick{$|+\rangle_a$} & \ctrl{-8} & \control{} & \ctrl{-7} & \ctrl{-6} & \ctrl{-5} & \ctrl{-4} & \ctrl{-3} & \ctrl{-2} & \control{} & \ctrl{-1} & \gate{H} & \meter{} \\
    \lstick{$|+\rangle_f$} & & \ctrl{-1} & \qw & & & & & & \ctrl{-1} & & \gate{H} & \meter{} \\
    \end{quantikz}
    };
    \end{tikzpicture}
    \caption{Circuit for measuring a stabilizer generator of the DK encoding. The top eight qubits are the data qubits. The auxiliary qubit $|+\rangle_a$ is read out to give the measurement result and the auxiliary qubit $|+\rangle_f$ is a flag qubit that is used to detect high-weight errors.}
    \label{fig:stab-circuit}
\end{figure}

To measure the stabilizers of the DK encoding, we use the standard Hadamard test circuit.
We also use a flag qubit~\cite{chao2018quantum} to detect when high-weight errors propagate to the data qubit; see \cref{fig:stab-circuit}.
The flag qubit detects all weight $\geq 2$ data qubit errors except for the stabilizer itself and errors that are stabilizer equivalent to weight $1$ errors.
However, a single fault on the auxiliary measurement qubit after all of the control gates can give an incorrect measurement result.
To mitigate this we can repeat the circuit twice.
We detect an error if either the flag qubit measurement outcome is $-1$ or the two stabilizer measurement outcomes disagree.

\subsection{State preparation}
\label{state-prep-ce}

First, we consider preparing Slater determinant states $|\psi_v\rangle$. 
Slater determinant states are common starting fermionic basis states for simulation purposes similar to qubit product states~\cite{kivlichan2018fermionswapnetwork,Jiang_2018}.
The convenient definition of Slater determinant states for our purposes is with $v \in \mathbb F_2^{|\mathcal L_v|}$ as follows~\cite{higgott2021optimal}
\begin{align} \label{eq:slater_v}
    & V_j |\psi_v\rangle = (-1)^{v_j} |\psi_v\rangle \quad j \in [L^2], \\
    & S_k |\psi_v\rangle = |\psi_v\rangle \quad k \in [L^2/2],
\end{align}
where $v_j$ is the $j$'th entry of $v$ and we recall that $S_k$ are the stabilizer generators of the encoding. 
In the case of the JW encoding there are no stabilizers and the Slater determinant states are simply computational basis states satisfying \cref{eq:slater_v}.
For the TT encoding, Slater determinant states are also represented by computational basis states, but using a parity mapping, where each binary vector $v$ corresponds to a certain parity sum of fermionic occupation numbers~\cite{harrison2024sierpinski}.
Our approach to preparing Slater determinant state in the DK encoding is analogous to the method given in~\cite{jiang2019majorana} for Majorana loop encodings.
We note however that the DK encoding has additional properties that allow for a more compact circuit.

As noted in~\cite{higgott2021optimal}, for the DK encoding the Slater determinant states factorize 
\begin{equation}
    |\psi_v\rangle = |v\rangle \otimes |\phi\rangle,
\end{equation}
where $|v\rangle \in (\mathbb C^2)^{\otimes |\mathcal L_v|}$ and $|\phi\rangle \in (\mathbb C^2)^{\otimes |\mathcal L_f^r|}$.
Let $R_k$ and $T_k$ denote the restriction of $S_k$ to the vertex qubits and to the face qubits, respectively. 
We have

\begin{equation}
\begin{split}
     R_k |v\rangle &= r_k |v\rangle, \quad T_k |\phi\rangle = t_k |\phi\rangle, \\ r_k&=t_k, \quad k \in [L^2/2].
\end{split}
\end{equation}

First, consider the case where $v = \mathbf 0$, i.e., the all-0 vector.
We can prepare the corresponding Slater determinant state by preparing all qubits in the $\ket 0$ state, measuring $T_k$ for $k \in [L^2/2]$ using auxiliary qubits, and post-selecting on the all $+1$ outcome case. 
However, as each of the $T_k$ measurements has a $50\%$ chance of giving outcome $+1$, the probability of success with this method is exponentially suppressed in the size of the system.

Let $\sigma \in \{ -1,+1 \}^{|\mathcal L_f^r|}$ denote the list of eigenvalues of the $S_k$, which we refer to as the syndrome in analogy with QEC codes.
Recall that any syndrome defines an equivalent subspace to the $\sigma = (1,1,\ldots,1)$ case, and so we do not actually need to post-select on the all $+1$ outcome case.
Instead, we need only to record the results of the $T_k$ measurements, which gives us $\sigma$ directly as $R_k |\mathbf 0\rangle = |\mathbf 0\rangle$ for all $j$.
If we measure the stabilizer generators again later in the circuit, we should compare their values to $\sigma$.

In addition, we must multiply some of the edge operators by $-1$ to ensure that all logical operators act correctly on the subspace labeled by $\sigma$.
The most efficient way to find the relevant edge operators is to use the destabilizers of the group generated by the $T_k$'s.
The destabilizers (or pure errors) of a stabilizer group with $m$ generators are a set of $m$ Pauli operators with the property that each operator anticommutes with exactly one of the stabilizer generators~\cite{aaronson2004improved}.
We can precompute the destabilizers using Gaussian elimination, and then for the observed syndrome $\sigma$ we construct a recovery operator as a product of destabilizers, where the destabilizer corresponding to $T_k$ is in the product iff $T_k$ had a $-1$ measurement outcome.
We then find all edge operators that anticommute with the recovery operator and multiply them by $-1$.

The case where $v \neq \mathbf 0$ is only slightly more complicated.
Here we calculate beforehand the eigenvalues of the $R_k$, $r_k|v\rangle = R_k |v\rangle$.
Next we measure the $T_k$ as before; let $\{t_k\}$ denote the measurement outcomes. 
Then the $k$'th entry of the syndrome is simply $r_k t_k$.

There is a further simplification of the circuit that we can make by preparing the face qubits differently.
Suppose that instead of preparing all face qubits in the $|0\rangle$ state we prepare the face qubits in even rows of the lattice in the $|+\rangle$ state and those in odd rows of the lattice in the $|+i\rangle$ state. 
With this change, the initial state is already in a defined $(\pm 1)$ eigenstate of the stabilizers centred at faces in the even rows, where the eigenvalue is simply given by $r_k$ for stabilizer $S_k$.
We therefore do not need to measure these stabilizer generators and so we only measure the $T_k$ centred at odd rows, computing the syndrome for these operators as before.
This simplification is akin to the state preparation technique for Calderbank-Shor-Steane (CSS) codes where the initial state is a $+1$ eigenstate of either the $Z$-type stabilizers or the $X$-type stabilizers~\cite{Nielsen_Chuang_2010}.
We comment that it may be possible to further reduce the number of stabilizer measurements required by starting from small entangled states rather than product states (as was demonstrated for the XY code~\cite{tsai2024mitigatingtemporal}).

We can also prepare other encoded states by a similar method.
Consider some subset of the edge operators of the DK encoding with non-overlapping support $\{E_{\ell m}\}$.
For each $\ell$ and $m$, we prepare the qubits in the support of $E_{\ell m}$ in a product state that is the $+1$ eigenstate of $E_{\ell m}$.
For example if $E_{\ell m} = X_\ell Y_m X_{f(\ell,m)}$ then we prepare the relevant qubits in the state $|+\rangle |+i\rangle |+\rangle$.
The second step is to measure the stabilizer generators, record the syndrome, and update the signs of edge operators as before.
We note that in this case we must measure the full set of stabilizer generators $\{ S_k \}$ and not the restricted operators $\{ T_j \}$.
This state preparation method generalizes to preparing $-1$ eigenstates of edge operators in the same way as the Slater determinant method.

\subsection{Logical operations}
\label{subsec:logical-operations}

In quantum error correction, logical operators map encoded states to encoded states.
The simplest set of logical operators for fermionic encodings, which is relevant to quantum simulation algorithms, is the set of arbitrary rotations of the form $e^{-i\theta{\tilde{F}}}$ where $\tilde{F}$ is a Pauli operator and a term in the encoded Hamiltonian. 

These operators are also the building blocks of Trotterized time evolution (the simplest method for implementing time evolution of a Hamiltonian on a quantum computer), which is defined as follows~\cite{lloyd1996universal,childs2021theory}:
\begin{equation}
e^{-iHt} \approx \prod_{n=0}^{t / T} \left(\rev{ \prod_{k} e^{-i  \tilde{F}_{k} (t/T)}} \right),
\end{equation}
where $H = \rev{\sum_{k} \tilde{F}_{k}}$ and $T$ is the number of Trotter steps.

The number of Trotter steps required for useful simulations of the Fermi-Hubbard model depends on the details of the simulation, but we can make a rough estimate using the first-order Trotter error bounds defined in~\cite{lin2022lecturenotesquantumalgorithms} as
\begin{equation}
    \epsilon = \left\| e^{-itH} - \left(e^{-i\frac{t}{L} H_1} e^{-i\frac{t}{L} H_2} \right)^T \right\|
    = \mathcal{O} \left( \frac{t^2}{T} \right).
\end{equation}
Therefore, to reach precision \( \epsilon \) in the operator norm, we need
\begin{equation}
    T = \mathcal{O}(t^2 \epsilon^{-1}).
\end{equation}
Roughly, this means that if we want to simulate Fermi-Hubbard evolution for time \(t=1\) at precision \(\epsilon=0.01\) then we need \(L=100\).
Or, with the same \(L=100\) we can simulate \(t=0.1\) with precision of \(\epsilon=10^{-4}\). 
We note that more precise bounds for the Fermi-Hubbard model depend on the magnitude of certain coefficients in the Hamiltonian~\cite{schubert2023trottererrorfermihubbard}.

\subsection{Readout}
\label{readout-subsec}

We first consider measuring the fermionic occupation number for each mode, which can be extracted by measuring all qubits in the $Z$ basis in each of the encodings we are considering.
Since fermionic systems in nature are global parity preserving, and computational basis qubit states encoded with all three encodings we consider retain that property, we can discard any runs where the total parity of the measurements is odd (assuming we start with an even number of fermions).
These measurements are therefore robust to any odd number of measurement errors. We call this type of postselection global parity postselection (GP).

The low weight of the vertex and edge operators in the DK encoding means that we can measure many of them simultaneously.
Consider a subset of vertex and edge operators with non-overlapping support. 
Any subset with this property can be measured simply by measuring the relevant qubits in the appropriate Pauli bases.
For example, to measure all of the vertex operators we simply measure all of the vertex qubits in the $Z$ basis.
For this measurement in particular, we can also extract a subset of the stabilizer eigenvalues.
Suppose we additionally measure the face qubits in even (odd) rows of the lattice in the $X$ ($Y$) basis.
Then by combining the relevant measurement outcomes, we can reconstruct the eigenvalues of the stabilizer generators associated with faces in even rows of the lattice.
We can then discard any runs with stabilizer measurement outcomes that do not match the outcomes observed in state preparation.
We call this readout technique \emph{stabilizer reconstruction} (SR).

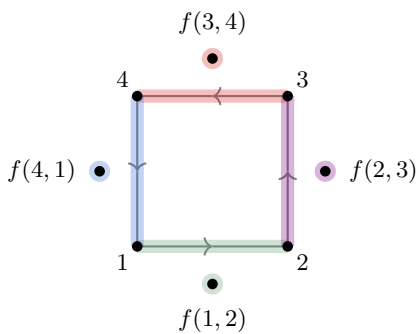
\begin{figure}[ht]
    \centering
\begin{tikzpicture}

\draw[thick,->,gray] (0,0) -- (1,0); %
\draw[thick,-,gray] (1,0) -- (2,0); %
\draw[line width=5pt,nicegreen,opacity=0.3] (0,0) -- (2,0); %

\draw[thick,->,gray] (2,0) -- (2,1); %
\draw[thick,-,gray] (2,1) -- (2,2); %
\draw[line width=5pt,nicepurple,opacity=0.3] (2,0) -- (2,2); %

\draw[thick,->,gray] (2,2) -- (1,2); %
\draw[thick,-,gray] (1,2) -- (0,2); %
\draw[line width=5pt,nicered,opacity=0.3] (2,2) -- (0,2); %

\draw[thick,->,gray] (0,2) -- (0,1); %
\draw[thick,-,gray] (0,1) -- (0,0); %
\draw[line width=5pt,niceblue,opacity=0.3] (0,2) -- (0,0); %

\fill[nicegreen,opacity=0.3] (1,-0.5) circle (4pt); %
\fill (1,-0.5) circle (2pt);
\node[anchor=north] at (1,-0.7) {$f(1,2)$};

\fill[nicepurple,opacity=0.3] (2.5,1) circle (4pt); %
\fill (2.5,1) circle (2pt);
\node[anchor=west] at (2.7,1) {$f(2,3)$};

\fill[nicered,opacity=0.3] (1,2.5) circle (4pt); %
\fill (1,2.5) circle (2pt);
\node[anchor=south] at (1,2.7) {$f(3,4)$};

\fill[niceblue,opacity=0.3] (-0.5,1) circle (4pt); %
\fill (-0.5,1) circle (2pt);
\node[anchor=east] at (-0.7,1) {$f(4,1)$};

\fill (0,0) circle (2pt);
\node[anchor=north east] at (0,0) {$1$};

\fill (2,0) circle (2pt);
\node[anchor=north west] at (2,0) {$2$};

\fill (2,2) circle (2pt);
\node[anchor=south west] at (2,2) {$3$};

\fill (0,2) circle (2pt);
\node[anchor=south east] at (0,2) {$4$};
\end{tikzpicture}
\caption{Efficient DK measurement schedule.
For each stabilizer in the lattice, we can measure the hopping term operators corresponding to the same color simultaneously. 
Each highlighted edge corresponds to two operators in the hopping term: $a_i^{\dagger} a_j$ and $a_j^{\dagger} a_i$; see \cref{eq:FH-CE-hop}.
}
\label{fig:ce-hopping-measurement}
\end{figure}

The low weight of the Hamiltonian terms in the DK encoding also allows for efficient extraction of the energy. In order to measure all the operators in the Fermi-Hubbard Hamiltonian, we only need to measure in nine different measurement bases. First, we can simultaneously extract the local occupation number operators $a_i^{\dagger} a_i$ and the Coulomb operators $a_i^{\dagger} a_i a_j^{\dagger} a_j$ by measuring the vertex qubits in the $Z$ basis. This is compatible with the fault-tolerant readout described above.
After that, we can simultaneously measure the non-overlapping parts of the hopping operators using eight sets of measurements as shown in~\cref{fig:ce-hopping-measurement}. 

In a square-lattice Fermi-Hubbard Hamiltonian encoded with the JW encoding, it is known how to extract the energy with 5 measurements using the procedure outlined by~\cite{cade2020neartermFH}. 
This procedure employs a simple, but non-Clifford, diagonalization circuit for the two parts of the hopping term. 
To the best of our knowledge, no efficient measurement procedure is known for ternary tree encoding.

\section{Numerical experiments} \label{sec:numerics}

\subsection{Stabilizer simulation background}

Stabilizer simulations are routinely used to benchmark the performance of quantum error-correcting codes with thousands of qubits, see e.g.~\cite{fowler2012surfacecodes,Grospellier2021combininghardsoft,vasmer2021cellular,breuckmann2022sinlgeshot}.
Stabilizer simulations can simulate logical operations as long as they are Clifford, and capture limited form of noise, namely stochastic Pauli channels.

\begin{figure}[ht]
    \centering

    \subfloat[Example decomposition of the DK encoding (half) hopping operator $e^{-i \theta X_i X_j Y_{f_{i,j}}}$ into 1- and 2-qubit gates.]{
        \centering
        \begin{tikzpicture}
            \node[scale=0.9] (circuit-decomp) {
                \begin{quantikz}
                 \lstick{1} & \gate{\sqrt{Y}} & \targ{} & \targ{} & \gate[style={draw=nicepurple}]{\textcolor{nicepurple}{R_z(\theta)}} & \targ{} & \targ{} & \gate{\sqrt{Y}} & \qw \\
                 \lstick{2} & \gate{\sqrt{Y}} & \ctrl{-1} &  & & & \ctrl{-1} & \gate{\sqrt{Y}} & \qw \\
                 \lstick{3} & \gate{\sqrt{X}} & & \ctrl{-2} & & \ctrl{-2} & & \gate{\sqrt{X}} & \qw \\
                \end{quantikz}
            };
        \end{tikzpicture}
        \label{fig:logical-decomp}
    }
    
    \subfloat[Clifford version of (a) corresponding to $e^{-i \frac{3\pi}{4}X_i X_j Y_{f_{i,j}}}$]{
        \centering
        
        \begin{tikzpicture}
            \node[scale=0.9] (circuit-decomp) {
                \begin{quantikz}
                 \lstick{1} & \gate{\sqrt{Y}} & \targ{} & \targ{} & \gate[style={draw=nicepurple}]{\textcolor{nicepurple}{S}} & \targ{} & \targ{} & \gate{\sqrt{Y}} & \qw \\
                 \lstick{2} & \gate{\sqrt{Y}} & \ctrl{-1} &  & & & \ctrl{-1} & \gate{\sqrt{Y}} & \qw \\
                 \lstick{3} & \gate{\sqrt{X}} & & \ctrl{-2} & & \ctrl{-2} & & \gate{\sqrt{X}} & \qw \\
                \end{quantikz}
            };
        \end{tikzpicture}
        
        \label{fig:clif-logical-decomp}
   }
   
   \subfloat[Comparison of the process fidelity of Clifford and non-Clifford logical rotations, calculated using \textsc{Qiskit}'s process tomography functionality~\cite{qiskit2024}.
   Each data point is the average value of five runs with 1000 shots each (the error bars are too small to be visible).]{
   \centering
    \includegraphics[width=\linewidth]{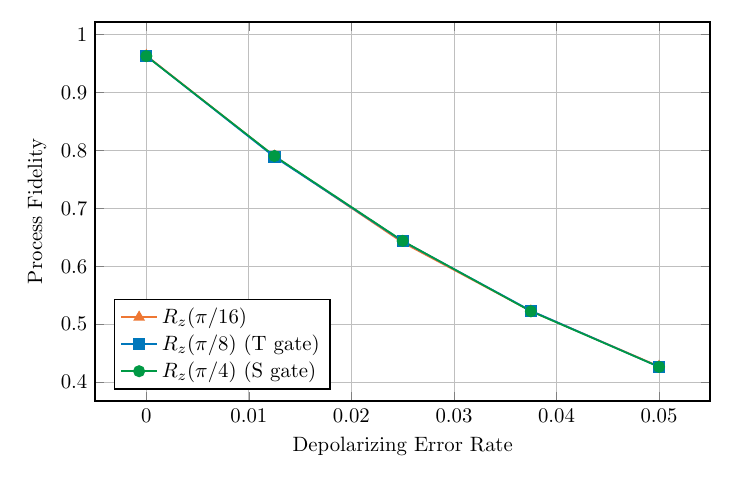}
    \label{fig:process-fid-plot}
   }
    \caption{Clifford and non-Clifford logical rotations.}
    \label{fig:logical-circuits}
\end{figure}

When using stabilizer simulations to benchmark the performance of fermionic encodings we also have the above limitations.
However, we argue that simulations in this restricted model can nevertheless serve as a reasonable proxy for performance.
First, note that the state preparation and readout circuits we discussed fall within the stabilizer formalism.
Requiring the logical operator circuits to be Clifford restricts the possible rotation angles we can choose in \cref{fig:logical-decomp} but we emphasize that the only part of these circuits that changes for different choices of rotation angle is one single-qubit gate, and therefore we would expect the Clifford circuits to accurately capture the error propagation in the generic circuits.
We also numerically confirm that the performance of such Clifford and non-Clifford circuits is identical in the stochastic Pauli error model via calculating process fidelities for varying error rates; see \cref{fig:process-fid-plot}.

The restriction to stochastic Pauli noise is perhaps the biggest limitation of our simulation, as this does not capture important device noise such as leakage~\cite{wood2018quantification,bermudez2019faulttolerant}, amplitude damping~\cite{myatt2000decoherence,turchette2000decoherence,chirolli2008decoherence}, or coherent errors.
Coherent errors in particular are damaging in deep circuits as they accumulate adversarially with increasing circuit depth.
However, we note that arbitrary Markovian errors (including coherent errors) can be tailored to stochastic Pauli errors using randomized compiling~\cite{wallman2016noise,hashim2021randomized}, and we expect our simulations to be especially predictive for circuits compiled in this way.

Our numerical results are enabled by \textsc{stim}~\cite{gidney2021stim}, a commonly used and powerful tool for testing QEC codes by constructing noisy Clifford circuits and sampling their outcomes under varying circuit-level error models. 
The advantages of using \textsc{stim} compared to prior approaches assessing noise resistance of fermionic encodings~\cite{chien2023simulating,hagge2023error} are: 1) the ability to simulate much larger sizes; 2) the flexibility to define hardware inspired error models; and 3) the requirement to construct explicit noisy Clifford circuits resembling circuits that one would run on a device.

The code is available at~\cite{Dyrenkova2025FermiStabilizers}. Data is available at~\cite{plot_data}.

\subsection{Error models}

We consider two error models: Standard Depolarizing (SD) and Superconducting Inspired (SI), which are described in \cref{tab:combined_tables}.
Variants of the SD model are often used as a benchmark in QEC research, while SI is meant to represent the errors that are common in superconducting devices and was itself inspired by~\cite{Gidney2022benchmarkingplanar}.
We do not consider idling errors in this work as adding idling errors would disadvantage encodings with high-weight logical operators, which are mostly applied sequentially in our simulations.

\begin{table}[ht]
    \centering
    \subfloat[Definitions of error types]{
        \begin{tabularx}{\linewidth}{|l|X|}
            \hline
            \textbf{Noise type} & \textbf{Definition} \\
            \hline
            $p_2$ & A 2-qubit depolarizing channel with rate $p_2$ is added after each 2-qubit Clifford gate. \\ 
            \hline
            $p_1$ & A single-qubit depolarizing channel with rate $p_1$ is added after each 1-qubit Clifford gate. \\
            \hline
            $p_{\mathrm{s}}$ & Initialization to $|0\rangle$ is followed by a bitflip channel with rate $p_{\mathrm{s}}$. \\
            \hline
            $p_{\mathrm{m}}$ & All measurements are done in the $Z$ basis and their results are flipped at a rate $p_{\mathrm{m}}$. \\
            \hline
        \end{tabularx}
        \label{tab:noisy_gates}
    }\\
     \subfloat[Comparison of noisy gatesets in SD and SI]{      
        \begin{tabularx}{\linewidth}{|X|X|X|}
            \hline
            \textbf{Name} & Standard Depolarizing & Superconducting Inspired \\
            \hline
            \textbf{Noisy Gateset} & $p_2=p$ & $p_2=p$ \\
            & $p_1=p$ &  $p_1=p/10$ \\
            & $p_{\mathrm{s}}=p$ & $p_{\mathrm{s}}=2p$ \\
            & $p_{\mathrm{m}}=p$ & $p_{\mathrm{m}}=5p$ \\
            \hline
        \end{tabularx}
        \label{tab:example_structure}
   }\\
   \subfloat[Error regimes that we simulated]{
        \begin{tabularx}{\linewidth}{|l|X|}
            \hline
            \textbf{Error Regime} & \textbf{Definition} \\
            \hline
            Aspirational & $p = 0.0001$, or 0.01\% \\
            \hline
            Intermediate & $p = 0.0005$, or 0.05\% \\
            \hline
            Near future & $p = 0.001$, or 0.1\% \\
            \hline
        \end{tabularx}
        \label{tab:error_regimes}
    }
    \caption{Error models}
    \label{tab:combined_tables}
\end{table}

\subsection{Simulated circuits}

\begin{figure*}[ht]
    \centering
    \scalebox{0.8}{
    \begin{tikzpicture}[
        box/.style = {draw, rectangle, minimum width=2cm, minimum height=1.2cm, thick, align=center},
        arrow/.style = {thick, -latex},
        line/.style = {thick}, %
        dashedline/.style = {dashed},
        dashedarrow/.style = {dashed, -latex},
        doubleline/.style = {thick, double, double distance=1pt},
        cloud/.style = {draw, rounded corners, minimum width=3cm, align=center, font=\small},
        dashedbox/.style = {draw, rectangle, dashed, thick, minimum width=2cm, minimum height=1.2cm, align=center},
        textstyle/.style = {font=\small, blue}
    ]

    \node (input) {$\left| 0 \right\rangle^{\otimes n}$};
    \node[box, right=1.0cm of input, align=center] (stateprep) {Eigenstate \\ prep};
    \node[dashedbox, right=1.0cm of stateprep, align=center] (encode) {Codespace \\ state prep};
    \node[box, right=1.0cm of encode, align=center] (circuit) {Logical circuit \\ \cref{fig:logical-circuits}};
    \node[dashedbox, right=1.0cm of circuit, align=center] (syndrome) {Syndrome \\ extraction};
    \node[box, right=1.0cm of syndrome, align=center] (measure) {Logical \\ measurement};

    \draw[arrow] (input) -- (stateprep);
    \draw[arrow] (stateprep) -- (encode);
    \draw[arrow] (encode) -- (circuit);
    \draw[arrow] (circuit) -- (syndrome);
    \draw[arrow] (syndrome) -- (measure);

    \node[cloud, below=1.0cm of encode, nicepurple] (nontrivial) {\textbf{Error is detected} \\ when syndrome is non-trivial};

    \draw[arrow] (encode.south) |- ++(0,-0.5) -| (nontrivial.north);
    \draw[arrow] (syndrome.south) |- ++(0,-0.5) -| (nontrivial.north);

    \node[cloud, right=1.0cm of measure, nicepurple] (obsError) {\textbf{Obs.\ error} \\ when different \\ from $\left| 0 \right\rangle^{\otimes n}$};

    \draw[doubleline] (measure.east) -- ++(0.5,0) -- ++(0.5,0) -- (obsError.west);

    \node[cloud, below=1.0cm of nontrivial, minimum width=3cm, align=center, nicepurple] (final) {\textbf{Undetected errors:} \\ triv. syndrome \\ with obs. error}; %

    \coordinate (merge) at ($(nontrivial.south)-(0,0.5)$);
    \coordinate (turn_logical) at ($(measure.south)-(0,0.5)$);
    
    \draw[line] (obsError.south) |- (merge); %
    \draw[line] (nontrivial.south) -- (merge); %
    \draw[dashedline] (measure.south) |- (turn_logical);
    \draw[dashedarrow] (turn_logical) -- ($(syndrome.south)-(0,0.5)$);
    \draw[arrow] (merge) -- (final.north);

    \end{tikzpicture}
    }
    \caption{\textbf{Simulated Circuit Structure.} Dashed elements are only present in some circuits. Codespace state preparation is only present in DK circuits. Syndrome extraction is only present in the DK circuits where non-destructive syndrome measurement is performed. Logical measurements contain information used for postselection in JW global parity postselection and stabilizer eigenvalue readout in DK described in \cref{readout-subsec}. Purple boxes indicate postselection steps from classical data.}
    
    \label{fig:simulated-circuits}
\end{figure*}

The overview of the structure of our circuits can be found \cref{fig:simulated-circuits}. 
The main idea is to create a noisy version of the circuit that implements
\begin{equation} \label{eq:mirror}
    U U^{\dagger} \lvert \psi_{\mathrm{init}} \rangle = \lvert \psi_{\mathrm{init}} \rangle.
\end{equation}
Applying the mirror circuit $UU^{\dagger}$ instead of just $U$ is the simplest way to make sure that the final logical measurement is deterministic without noise, which is a requirement for \textsc{stim} simulations. 
Therefore, $\lvert\psi_{\mathrm{init}} \rangle$ is chosen to be an eigenstate of the final measurement basis. 
For simplicity, in our simulations $\lvert\psi_{\mathrm{init}} \rangle$ is always the +1 eigenstate of the measured basis, but that is not a requirement and does not affect the outcomes (\rev{choosing a different initial eigenstate would just imply a corresponding update of the expected logical operator values}). 

\rev{We note that, in principle, \textsc{stim} could also be used to simulate a non-mirror circuit $U$. However, we do not expect this to change our conclusions. The logical error rate is determined solely by the residual Pauli error acting on the final state—not by the structure or support of the ideal output state—and all gates in the mirror circuit are noisy. Thus, replacing a mirror circuit with a corresponding non-mirror circuit of the same depth would not qualitatively affect the resulting logical error rates.}

\textit{Eigenstate preparation} in \cref{fig:simulated-circuits} refers to single qubit rotations that put physical qubits into the +1 eigenstate of the final logical measurement.
\textit{Codespace state preparation} refers to the methods described in \cref{state-prep-ce}.
\textit{Syndrome extraction} is achieved through non-destructive stabilizer measurement circuit \cref{fig:stab-circuit}. Depending on the error mitigation method, the syndrome could be a combination of measurement results from codespace state preparation, syndrome extraction and logical measurement.
The \textit{logical circuit} block in \cref{fig:simulated-circuits} refers to $UU^\dagger$ in \cref{eq:mirror}, where $U$ represents a logical computation performed with an encoding ($U^{\dagger}$ is automatically generated with \textsc{stim} and is always counted towards the total depth of the circuit).
In our simulations, we consider two example logical computations, which we refer to as random logical circuits and full Trotter circuits.
For random logical circuits we apply some number of rotations of Hamiltonian terms (generated automatically using \textsc{Cirq}~\cite{cirq}) in random order by drawing terms uniformly at random from the Hamiltonian without replacement~\cite{chien2023simulating}. 
\rev{This is reminiscent of the idea of "randomized product formulas" ~\cite{Childs2019fasterquantum}.}
For full Trotter circuits, we apply all Hamiltonian terms some number of times.
In this case, there is an optimized circuit for the JW encoding utilizing fermionic swap networks~\cite{kivlichan2018fermionswapnetwork,stanisic2022observing,hagge2022optimal}, which we can simulate using \textsc{stim} as the fermionic swap operation is a Clifford operator.
For the DK encoding, we obtain an efficient circuit by applying the Hamiltonian terms in an order that maximizes parallel operations.
No similar optimizations are currently available for the TT encoding to the best of our knowledge.
For completeness, we provide tables of one- and two-qubit gate counts for all cases in \cref{app:depth-tables}.

\subsection{Sampling}\label{subsec:sampling}

For each quantum circuit, a sample is represented by two binary vectors: the syndrome vector \( \mathbf{s} \) and the observables vector \( \mathbf{o} \). The syndrome vector \( \mathbf{s} = (s_1, s_2, \dots, s_n) \), where \( s_i \in \{0, 1\} \), encodes the detection of errors in the system. 
The error detection rate is defined as 
\begin{equation}
R_{\mathrm{det}} = \frac{n_{\mathrm{discard}}}{n_{\mathrm{samp}}},
\end{equation}
where \( n_{\mathrm{samp}} \) is the total number of samples in the simulation and $n_{\mathrm{discard}}$ is the number of samples where $\mathbf s$ has at least one nonzero entry.

After postselection, the total number of samples used for calculating observable error rates is denoted as \( n_{\mathrm{samp,post}} \):
\begin{equation}
n_{\mathrm{samp,post}} = n_{\mathrm{samp}} - n_{\mathrm{discard}}.
\end{equation}

The observables vector \( \mathbf{o} = (o_1, o_2, \dots, o_m) \), where \( o_j \in \{0, 1\} \), tracks the outcomes of measured observables. A sample is counted toward the \( R_{\mathrm{obs,any}} \) if \( \mathbf{o} \) contains at least one non-zero entry,
and this rate is calculated as 
\begin{equation}
R_{\mathrm{obs,any}} = \frac{\text{Number of samples with any } o_j = 1}{n_{\mathrm{samp,post}}}.
\end{equation}
Additionally, the worst local observable rate \( R_{\mathrm{obs,worst}} \) is determined as the maximum local observable rate, where the local observable rate for observable \( j \) is defined as 
\begin{equation}
R_{\mathrm{obs,local}}^{[j]} = \frac{\text{Number of samples with } o_j = 1}{n_{\mathrm{samp,post}}}
\end{equation}
and 
\begin{equation}
R_{\mathrm{obs,worst}} = \max_{j} (R_{\mathrm{obs,local}}^{[j]}).
\end{equation}

In all simulations, \( n_{\mathrm{samp}} = 100{,}000 \), except for the VQED comparison, where \( n_{\mathrm{samp}} = 10{,}000 \).

\subsection{Results}

\begin{figure*}
    \centering
        \subfloat[Standard depolarizing error model at different error rates]{
        \centering
        \includegraphics[width=0.95\linewidth]{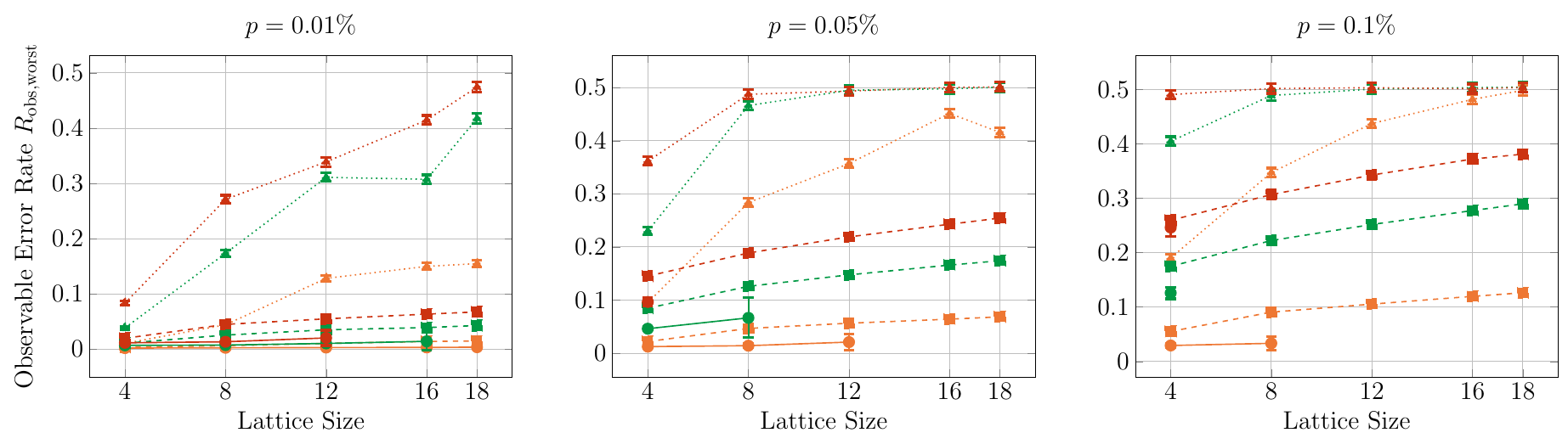}
        
    \label{fig:enc-comparison-a}
    }

   \subfloat[Standard depolarizing error model at p=$0.01\%$]{
        \centering
        \includegraphics[width=0.95\linewidth]{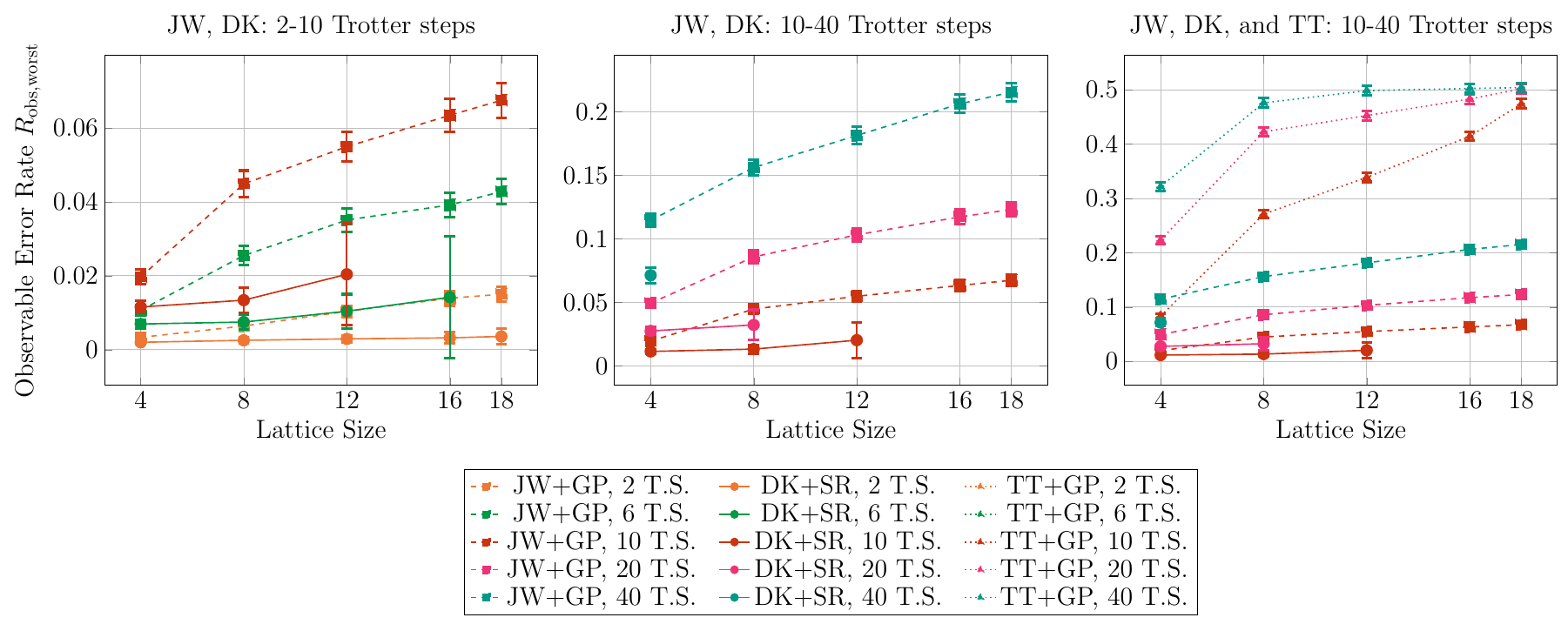}
        \label{fig:enc-comparison-b}
        }
    \caption{
    \textbf{Performance of fermionic encodings in the standard depolarizing error model.}
    \rev{We consider the DK, JW, and TT encodings, with stabilizer reconstruction (SR) and global parity postselection (GP).}
    One Trotter step \rev{(T.S.)} corresponds to an application of all the Hamiltonian terms, and we measure the occupation number at the end of the circuit.
    $R_{\mathrm{obs,worst}}$ is the maximum local observable error rate and data points with error detection rate $R_{\text{det}}\geq0.995$ are excluded.
    Error bars are computed using bootstrap resampling with 1000 resamples.
    (a) compares encoding performance across error rates.
    (b) zooms in on the lowest error rate, $p=0.01\%$: the left plot magnifies DK and JW data from (a), the center compares deeper circuits for DK and JW, and the right includes all three encodings (note the y-axis scale change).
    }
    \label{fig:enc-comparison}
\end{figure*}

\begin{figure*}
    \centering
   \subfloat[Observable errors in occupation number basis at $p=0.05\%$ (standard depolarizing error model)]{
        \centering
        \includegraphics[width=0.85\linewidth]{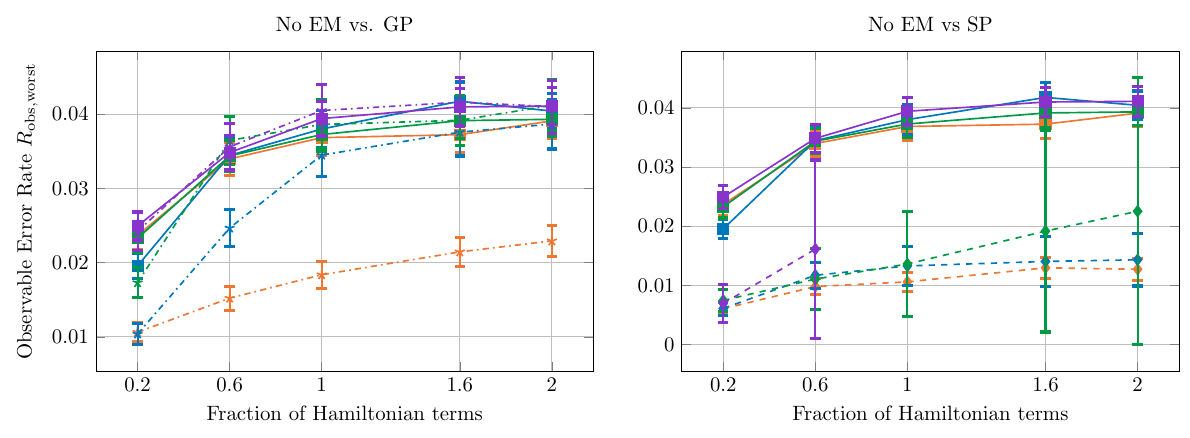}
        \label{fig:compact-obs-occ}
    }
    
    \subfloat[Observable errors in hopping operator basis at $p=0.05\%$ (standard depolarizing error model)]{
        \centering
        \includegraphics[width=0.85\linewidth]{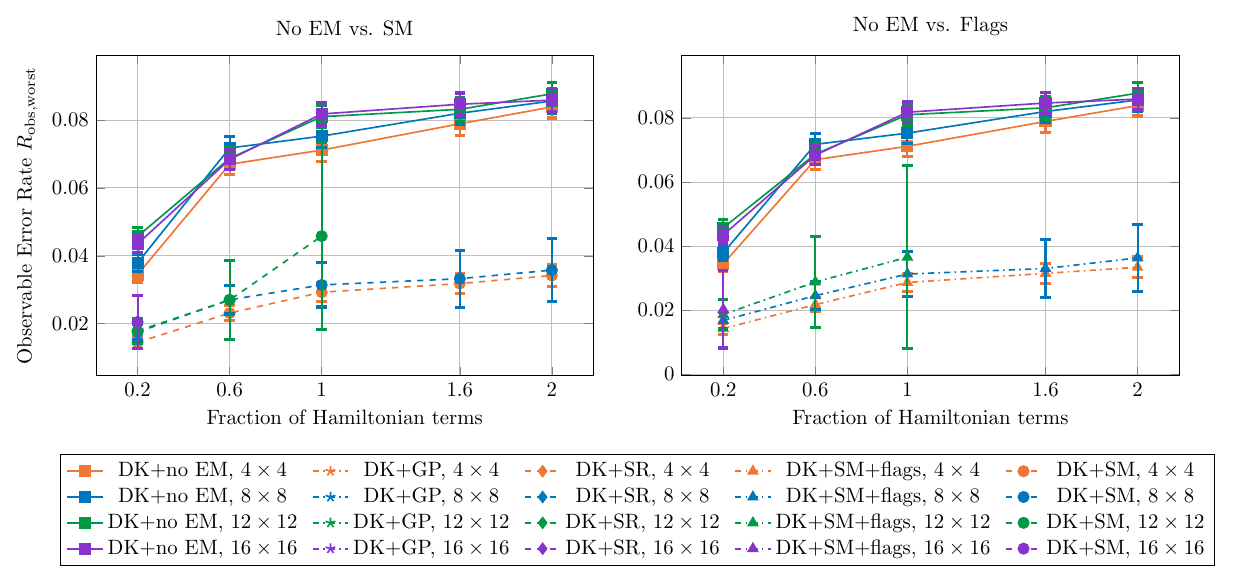}
        \label{fig:compact-obs-hop}
    }
    
   \subfloat[Error detection rates at different physical error rates (standard depolarizing error model)]{
        \centering
        \includegraphics[width=0.95\linewidth]{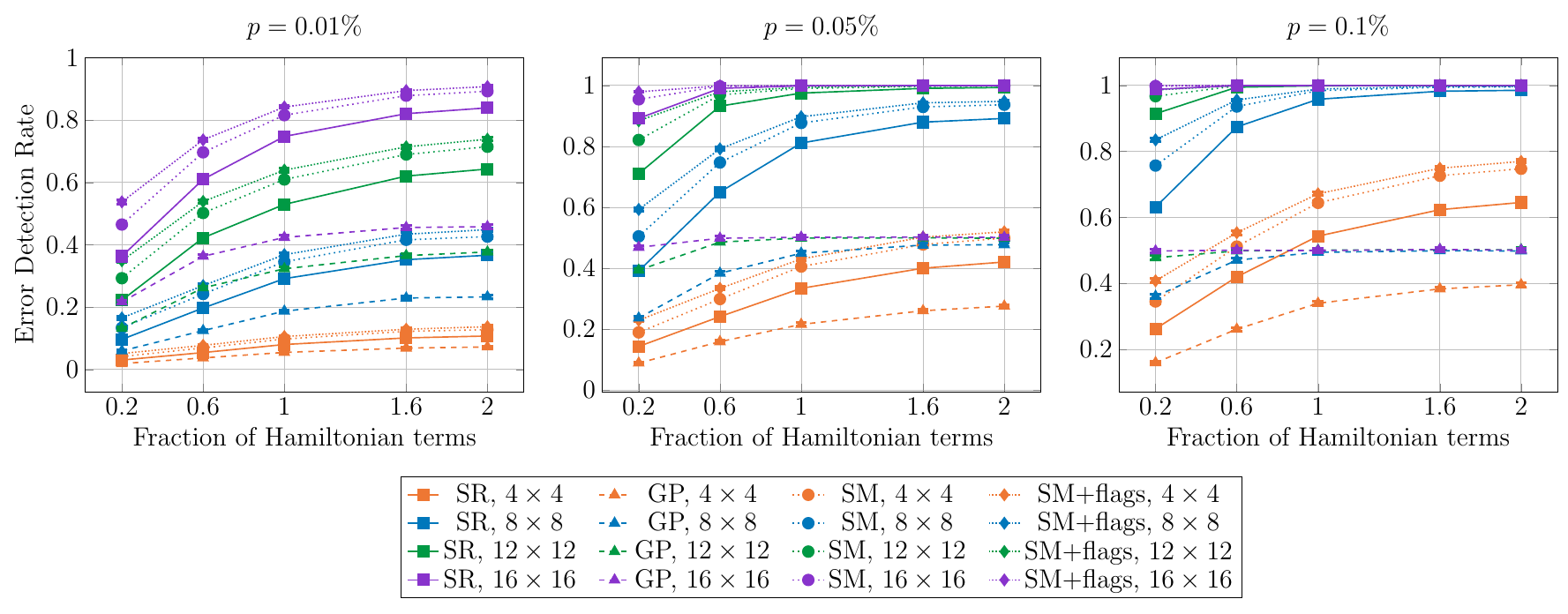}
        \label{fig:compact-det}
        }

  \caption{
  \textbf{Performance of various postselection methods in the DK encoding (SD error model).} 
  Each circuit applies a random sequence of logical operators, with operator count given as a fraction of Hamiltonian terms. \rev{The fraction of Hamiltonian terms takes on values up to $2$ because we include the mirror circuit depth.}
  Error bars are computed using bootstrap resampling with 1000 resamples.
  (a) and (b) showcase improvements in accuracy enabled by global parity postselection (GP) and stabilizer reconstruction (SR) for occupation number measurements, as well as stabilizer measurement (SM) and stabilizer measurement with flag qubits (SM+flags) for hopping term measurements, compared with no error mitigation (no EM). 
  Data points with error detection rate $R_{\mathrm{det}} \geq 0.995$ are excluded.
  The legend in (b) is shared by (a) and (b).
  (c) presents a comparison of the error detection rates of different postselection methods for varying error rates. 
  }
    \label{fig:compact-comparison}
\end{figure*}

First, we consider the performance of the JW, DK, and TT encodings with full Trotter circuits and occupation number measurements. 
This means that the observable vector $\mathbf{o} = (o_1, o_2, \dots, o_m)$ (see~\cref{subsec:sampling}) records the outcomes of computational basis ($Z$-basis) measurements on each qubit (for DK, this includes only vertex qubits).
For JW, we implement fermionic swap network circuits (\cref{subsec:logical-operations}) and global parity postselection (\cref{readout-subsec}). For DK, we implement stabilizer reconstruction as described in~\cref{readout-subsec}. We compare these to unoptimized, TT encoding \rev{with global parity postselection}, which is an asymptotically optimal one-to-one encoding. From~\cref{fig:enc-comparison}, it is evident that the optimized JW and DK circuits produce less noisy results than TT. The winner between JW and DK is less clear. While DK with stabilizer reconstruction (SR) shows a clear accuracy improvement, \rev{as well as almost constant undetected error rates with respect to lattice size,} this method quickly exhausts the sample budget $n_{\mathrm{samp}} = 100{,}000$ with increasing circuit depth, \rev{lattice sizes,} and physical error rate. This is reflected in the incompleteness of DK curves, as we include only data points calculated from at least 500 samples. \rev{The dramatic increase of variance in some curves is another visible artifact of running out of sample budget.} The most promising results obtained with our naive DK circuit construction were 10 Trotter steps for a $12 \times 12$ lattice or 6 Trotter steps for a $16 \times 16$ lattice at the lowest physical error rate of $p = 0.01\%$. But in that regime, the accuracy returns compared to JW are diminishing.

Next, we analyze the performance of various postselection methods for the DK encoding in~\cref{fig:compact-comparison}. We simulate random logical circuits and both occupation number and hopping term measurements. For the hopping term measurements, the observable vector $\mathbf{o}$ corresponds to measurement results of a multi-qubit observable $X_{i} X_{j} X_{f_{(i,j)}}$, arbitrarily chosen from the four possible encoded hopping operators listed in~\cref{eq:FH-CE-hop}.
Since we estimate the hopping term expectation by measuring all four separately, stabilizers must be measured non-destructively to obtain the syndrome vector $\mathbf{s}$ (see~\cref{subsec:stab-meas}). 
We choose an error rate of $p = 0.05\%$ in~\cref{fig:compact-obs-occ,fig:compact-obs-hop} to illustrate the tradeoff between accuracy and sampling requirements for different methods.
The two plots in~\cref{fig:compact-obs-occ} highlight the advantage of stabilizer reconstruction over global parity postselection in the DK encoding. Accuracy is improved due to the finer-grained error detection enabled by SR. This is also evident in~\cref{fig:compact-det}, where GP saturates at $R_{\text{det}}=0.5$ because a fully random occupation number measurement yields even global parity half of the time and odd global parity the other half.
The two plots in~\cref{fig:compact-obs-hop} compare stabilizer measurement postselection with and without flag qubits in the hopping term measurement basis. 
We observe little difference between flagged and non-flagged stabilizer measurements in both observable error rates and error detection rates (see~\cref{fig:compact-det}).

In \cref{app:additional-figures}, we provide additional results for the superconducting inspired error model, hopping term measurements, GP with the JW encoding. \rev{We also include a feasibility study for the and virtual quantum error detection (VQED) with the DK encoding in \cref{app:vqed}}.

\section{Conclusion} \label{sec:conclusion}

In this work we used modern stabilizer simulation tools to estimate the performance of fermionic encodings on near-term devices.
We were able to assess the performance of encodings in regimes that are potentially impractical for classical computation~\cite{agrawal2024quantifyingfaulttolerantsimulation, papic2025neartermfermionicsimulationsubspace}, reaching lattice sizes of up to $18 \times 18$ and up to 40 Trotter steps. 
To achieve this our error model was necessarily restricted to stochastic Pauli noise, but realistic noise in devices can be tailored to this form at little or no experimental cost.

We note that the limiting factor in our simulations was the performance of the encodings, rather than the computational resources required. 
The most computationally demanding aspect of our simulations was circuit generation, which could be further optimized. 
We utilized high-performance computing resources to test various combinations of encodings, circuits, and error models. 
However, we emphasize that results for a single set of simulation parameters can be efficiently obtained on a standard laptop.

In our simulations, we observed that the optimal average operator weight of the TT encoding cannot compete with depth reductions provided with specialized methods such as the JW encoding with fermionic swap netwoks and the DK encoding with error mitigation. 
We also observed that although the DK encoding with error detection provides an accuracy improvement, the returns are diminishing due to the high sampling requirements.
The most promising regime for the DK encoding is the aspirational regime ($p=0.01\%$).
However, even in this case the DK encoding exceeds our sample budget of 100,000 shots at 10 Trotter steps for the $12\times12$ lattice or 6 Trotter steps for the $16\times16$ lattice.
Other techniques we tested such as flag qubits and virtual quantum error detection (VQED) did not show improvements in accuracy or sampling requirements.
Furthermore, in the aspirational regime, the JW encoding with fermionic swap networks and global parity postselection also provides relatively low-error results. 

Overall, our results indicate that reductions in operator locality and error mitigation enabled by the DK encoding are not enough to enable significant improvements in near-term condensed matter simulation involving Trotterization.
However, recent work~\cite{nigmatullin2024experimentaldemonstrationbreakevencompact} gives more optimized circuits and improved error detection techniques for the DK encoding, which we would expect to improve our results.
Considering the superior performance of the DK encoding in the aspirational regime, we could consider concatenating the DK encoding with a quantum error-correcting code such as the surface code.
The surface code layer could be used to reduce the logical error rate sufficiently so that the DK encoding postselection would be effective.
This combination could have superior resource requirements to either encoding on its own, and is compatible with stabilizer simulation.

Another next step for our work could be to test the validity of our simulations by implementing our circuits on quantum hardware, where the correct choice of encoding for a particular device will depend both on the characteristics of the device and the nature of the application.
For example, our circuits assume auxiliary qubits that can couple to eight data qubits, which is not possible on, e.g.,\ IBM's devices with heavy hex connectivity.
We may therefore need to consider compact encodings defined on other lattices~\cite{derby2021compact} in this case.
But our simulation tools are flexible and could be used to estimate the performance of other encodings with minor modifications. 
After selecting a device and application, it would be interesting to compare stabilizer simulations with device performance, both with and without randomized compiling. 
We expect randomized compiling to improve agreement between simulations and experiments, as well as enhance the performance of fermionic encodings with error detection.
While achieving quantum advantage with fermionic encodings in the near term is uncertain, stabilizer simulations will remain crucial for assessing quantum circuit performance at the scale where quantum advantage is possible.

\begin{acknowledgments}

The authors thank Toby Cubitt, Joel Klassen, Nathan Wiebe, and the Whitfield Group for helpful discussions. Research at Perimeter Institute is supported in part by
the Government of Canada through the Department of
Innovation, Science and Economic Development Canada
and by the Province of Ontario through the Ministry of
Colleges and Universities.
We acknowledge the support of the Natural Sciences and Engineering Research Council of Canada (NSERC).
This research was enabled in part by support provided by Compute Ontario (\url{https://computeontario.ca}) and the Digital Research Alliance of Canada (\url{https://alliancecan.ca}).
R.L. and E.D. thank Mike and Ophelia Lazaridis for funding.

\end{acknowledgments}

\clearpage

\appendix

\section{Depth tables}
\label{app:depth-tables}

We provide one-qubit (1Q) and two-qubit (2Q) gate counts in circuits that were used in our plots in \cref{tab:depth_table_dk_sr_fractions,tab:depth_table_dk_sm_fractions,tab:depth_table_dk_sr_trotter,tab:depth_table_jw_gp_trotter,tab:depth_table_ter_trotter}. 
This provides a rough prediction for how many noisy gates a simulation with different encodings can tolerate according to our results. 
In some sense, this is an upper bound on the number of gates that can be performed in similar implementations; if the Clifford simulation shows failure to extract information after a certain number of gates, then it is unlikely that the same circuit would perform better in a real device.

\begin{table*}
\begin{tabular}{|c|c|c|c|c|c|c|c|c|c|c|c|c|}
\hline
Fraction of Hamiltonian terms & \multicolumn{3}{|c|}{Lattice size $4\times4$} & \multicolumn{3}{|c|}{Lattice size $8\times8$} & \multicolumn{3}{|c|}{Lattice size $12\times12$} & \multicolumn{3}{|c|}{Lattice size $16\times16$} \\
\hline
  & 2Q & 1Q & Total & 2Q & 1Q & Total & 2Q & 1Q & Total & 2Q & 1Q & Total \\
\hline
0.2 & 96 & 192 & 288 & 296 & 622 & 918 & 664 & 1450 & 2114 & 1226 & 2698 & 3924 \\
\hline
0.6 & 218 & 390 & 608 & 710 & 1420 & 2130 & 1792 & 3762 & 5554 & 3154 & 6542 & 9696 \\
\hline
1.0 & 334 & 624 & 958 & 1216 & 2330 & 3546 & 2510 & 5304 & 7814 & 4562 & 9392 & 13954 \\
\hline
1.6 & 430 & 782 & 1212 & 1596 & 3100 & 4696 & 3396 & 6724 & 10120 & 5962 & 12084 & 18046 \\
\hline
2.0 & 466 & 830 & 1296 & 1604 & 3180 & 4784 & 3566 & 7096 & 10662 & 6310 & 12592 & 18902 \\
\hline
\end{tabular}
\caption{DK Encoding, occupation number measurement, stabilizer reconstruction error detection, random logical circuits.}
\label{tab:depth_table_dk_sr_fractions}
\end{table*}

\begin{table*}
\begin{tabular}{|c|c|c|c|c|c|c|c|c|c|c|c|c|}
\hline
Fraction of Hamiltonian terms & \multicolumn{3}{|c|}{Lattice size $4\times4$} & \multicolumn{3}{|c|}{Lattice size $8\times8$} & \multicolumn{3}{|c|}{Lattice size $12\times12$} & \multicolumn{3}{|c|}{Lattice size $16\times16$} \\
\hline
  & 2Q & 1Q & Total & 2Q & 1Q  & Total & 2Q & 1Q & Total & 2Q & 1Q & Total \\
\hline
0.2 & 208 & 232 & 440 & 746 & 792 & 1538 & 1704 & 1972 & 3676 & 2868 & 3142 & 6010 \\
\hline
0.6 & 330 & 446 & 776 & 1186 & 1658 & 2844 & 2752 & 4130 & 6882 & 4790 & 7002 & 11792 \\
\hline
1.0 & 446 & 662 & 1108 & 1602 & 2528 & 4130 & 3634 & 5720 & 9354 & 6374 & 10154 & 16528 \\
\hline
1.6 & 540 & 846 & 1386 & 2044 & 3292 & 5336 & 4414 & 7138 & 11552 & 7696 & 12674 & 20370 \\
\hline
2.0 & 566 & 854 & 1420 & 2054 & 3358 & 5412 & 4652 & 7612 & 12264 & 8178 & 13406 & 21584 \\
\hline
\end{tabular}
\caption{DK Encoding, hopping term measurement, stabilizer measurement error detection, random logical circuits.}
\label{tab:depth_table_dk_sm_fractions}
\end{table*}

\begin{table*}
\begin{tabular}{|c|c|c|c|c|c|c|c|c|c|c|c|c|c|c|c|}
\hline
Trotter steps & \multicolumn{3}{|c|}{Lattice size $4\times4$} & \multicolumn{3}{|c|}{Lattice size $8\times8$} & \multicolumn{3}{|c|}{Lattice size $12\times12$} & \multicolumn{3}{|c|}{Lattice size $16\times16$} & \multicolumn{3}{|c|}{Lattice size $18\times18$} \\
\hline
  & 2Q & 1Q & Total & 2Q & 1Q & Total & 2Q & 1Q & Total & 2Q & 1Q & Total & 2Q & 1Q & Total \\
\hline
2 & 242 & 385 & 627 & 778 & 1439 & 2217 & 1610 & 3089 & 4699 & 2730 & 5313 & 8043 & 3398 & 6641 & 10039 \\
\hline
4 & 470 & 753 & 1223 & 1490 & 2787 & 4277 & 3064 & 5941 & 9005 & 5188 & 10181 & 15369 & 6450 & 12711 & 19161 \\
\hline
6 & 698 & 1121 & 1819 & 2202 & 4135 & 6337 & 4518 & 8793 & 13311 & 7632 & 15029 & 22661 & 9498 & 18765 & 28263 \\
\hline
8 & 926 & 1489 & 2415 & 2914 & 5483 & 8397 & 5972 & 11645 & 17617 & 10086 & 19887 & 29973 & 12546 & 24825 & 37371 \\
\hline
10 & 1154 & 1857 & 3011 & 3626 & 6831 & 10457 & 7426 & 14497 & 21923 & 12530 & 24735 & 37265 & 15594 & 30879 & 46473 \\
\hline
20 & 2294 & 3697 & 5991 & 7186 & 13571 & 20757 & 14696 & 28757 & 43453 & 24780 & 49005 & 73785 & 30834 & 61167 & 92001 \\
\hline
40 & 4574 & 7377 & 11951 & 14306 & 27051 & 41357 & 29236 & 57277 & 86513 & 49270 & 97535 & 146805 & 61314 & 121737 & 183051 \\
\hline
\end{tabular}
\caption{DK Encoding, occupation number measurement, stabilizer reconstruction and full Trotter circuits.}
\label{tab:depth_table_dk_sr_trotter}
\end{table*}

\begin{table*}
\begin{tabular}{|c|c|c|c|c|c|c|c|c|c|c|c|c|c|c|c|}
\hline
Trotter steps & \multicolumn{3}{|c|}{Lattice size $4\times4$} & \multicolumn{3}{|c|}{Lattice size $8\times8$} & \multicolumn{3}{|c|}{Lattice size $12\times12$} & \multicolumn{3}{|c|}{Lattice size $16\times16$} & \multicolumn{3}{|c|}{Lattice size $18\times18$} \\
\hline
  & 2Q & 1Q & Total & 2Q & 1Q & Total & 2Q & 1Q & Total & 2Q & 1Q & Total & 2Q & 1Q & Total \\
\hline
2 & 140 & 117 & 257 & 642 & 419 & 1061 & 1436 & 919 & 2355 & 2562 & 1611 & 4173 & 3272 & 2029 & 5301 \\
\hline
4 & 282 & 241 & 523 & 1478 & 977 & 2455 & 3258 & 2275 & 5533 & 5780 & 4071 & 9851 & 7364 & 5155 & 12519 \\
\hline
6 & 424 & 365 & 789 & 2318 & 1541 & 3859 & 5086 & 3639 & 8725 & 9006 & 6545 & 15551 & 11464 & 8299 & 19763 \\
\hline
8 & 566 & 489 & 1055 & 3158 & 2105 & 5263 & 6914 & 5003 & 11917 & 12232 & 9019 & 21251 & 15564 & 11443 & 27007 \\
\hline
10 & 708 & 613 & 1321 & 3998 & 2669 & 6667 & 8742 & 6367 & 15109 & 15458 & 11493 & 26951 & 19664 & 14587 & 34251 \\
\hline
20 & 1418 & 1233 & 2651 & 8198 & 5489 & 13687 & 17882 & 13187 & 31069 & 31588 & 23863 & 55451 & 40164 & 30307 & 70471 \\
\hline
40 & 2838 & 2473 & 5311 & 16598 & 11129 & 27727 & 36162 & 26827 & 62989 & 63848 & 48603 & 112451 & 81164 & 61747 & 142911 \\
\hline
\end{tabular}
\caption{JW Encoding, occupation number measurement, global parity postselection, full Trotter circuits.}
\label{tab:depth_table_jw_gp_trotter}
\end{table*}

\begin{table*}
\begin{tabular}{|c|c|c|c|c|c|c|c|c|c|c|c|c|c|c|c|}
\hline
Trotter steps & \multicolumn{3}{|c|}{Lattice size $4\times4$} & \multicolumn{3}{|c|}{Lattice size $8\times8$} & \multicolumn{3}{|c|}{Lattice size $12\times12$} & \multicolumn{3}{|c|}{Lattice size $16\times16$} & \multicolumn{3}{|c|}{Lattice size $18\times18$} \\
\hline
  & 2Q & 1Q & Total & 2Q & 1Q & Total & 2Q & 1Q & Total & 2Q & 1Q & Total & 2Q & 1Q & Total \\
\hline
2 & 872 & 664 & 1536 & 4658 & 3740 & 8398 & 12060 & 9152 & 21212 & 23210 & 17676 & 40886 & 31158 & 23656 & 54814 \\
\hline
4 & 1704 & 1384 & 3088 & 8510 & 7450 & 15960 & 22396 & 18314 & 40710 & 41376 & 35300 & 76676 & 53798 & 46970 & 100768 \\
\hline
6 & 2536 & 2072 & 4608 & 12294 & 11194 & 23488 & 32726 & 27468 & 60194 & 59670 & 53164 & 112834 & 76672 & 70856 & 147528 \\
\hline
8 & 3396 & 2728 & 6124 & 16146 & 14854 & 31000 & 42846 & 36480 & 79326 & 77848 & 70890 & 148738 & 99166 & 93942 & 193108 \\
\hline
10 & 4240 & 3410 & 7650 & 19914 & 18688 & 38602 & 53244 & 45726 & 98970 & 95898 & 87952 & 183850 & 121712 & 117366 & 239078 \\
\hline
20 & 8484 & 6654 & 15138 & 38652 & 37450 & 76102 & 104464 & 91002 & 195466 & 186002 & 177196 & 363198 & 234404 & 234538 & 468942 \\
\hline
40 & 16944 & 13314 & 30258 & 76432 & 74850 & 151282 & 207134 & 182432 & 389566 & 369880 & 354218 & 724098 & 460258 & 470880 & 931138 \\
\hline
\end{tabular}
\caption{TT Encoding, occupation number measurement, full Trotter circuits.}
\label{tab:depth_table_ter_trotter}
\end{table*}

\section{Additional simulation data}
\label{app:additional-figures}

\begin{figure*}
    \centering
        \subfloat[Superconducting inspired error model at different error rates]{
        \centering
        \includegraphics[width=0.95\linewidth]{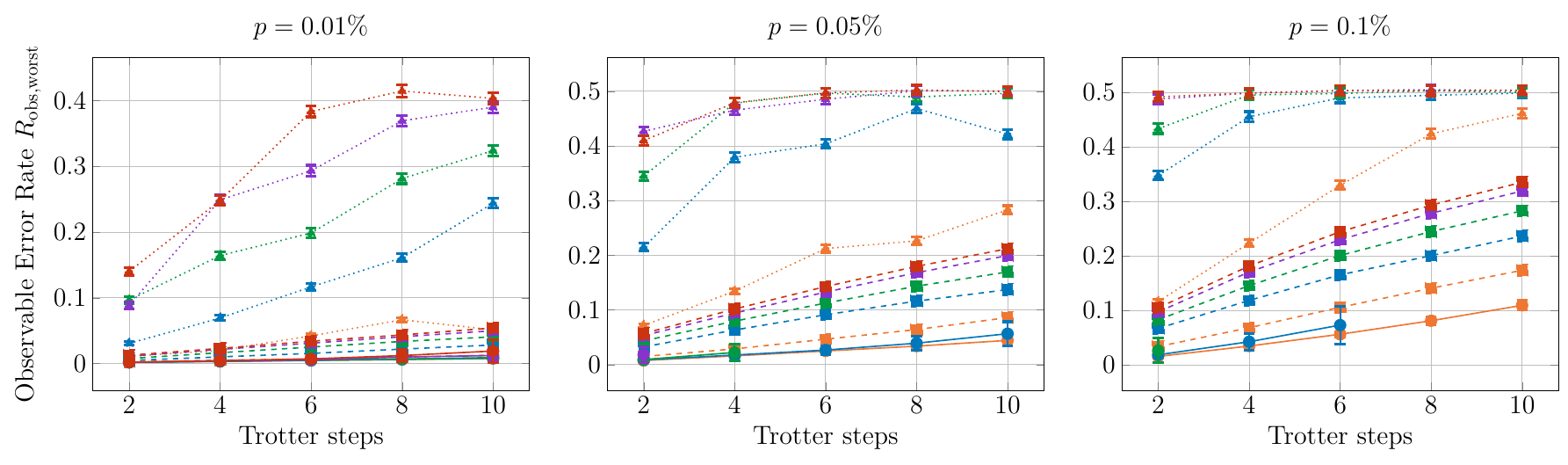}
        
    \label{fig:enc-comparison-si-a}
    }

   \subfloat[Superconducting inspired error model at p=$0.01\%$]{
        \centering
        \includegraphics[width=0.95\linewidth]{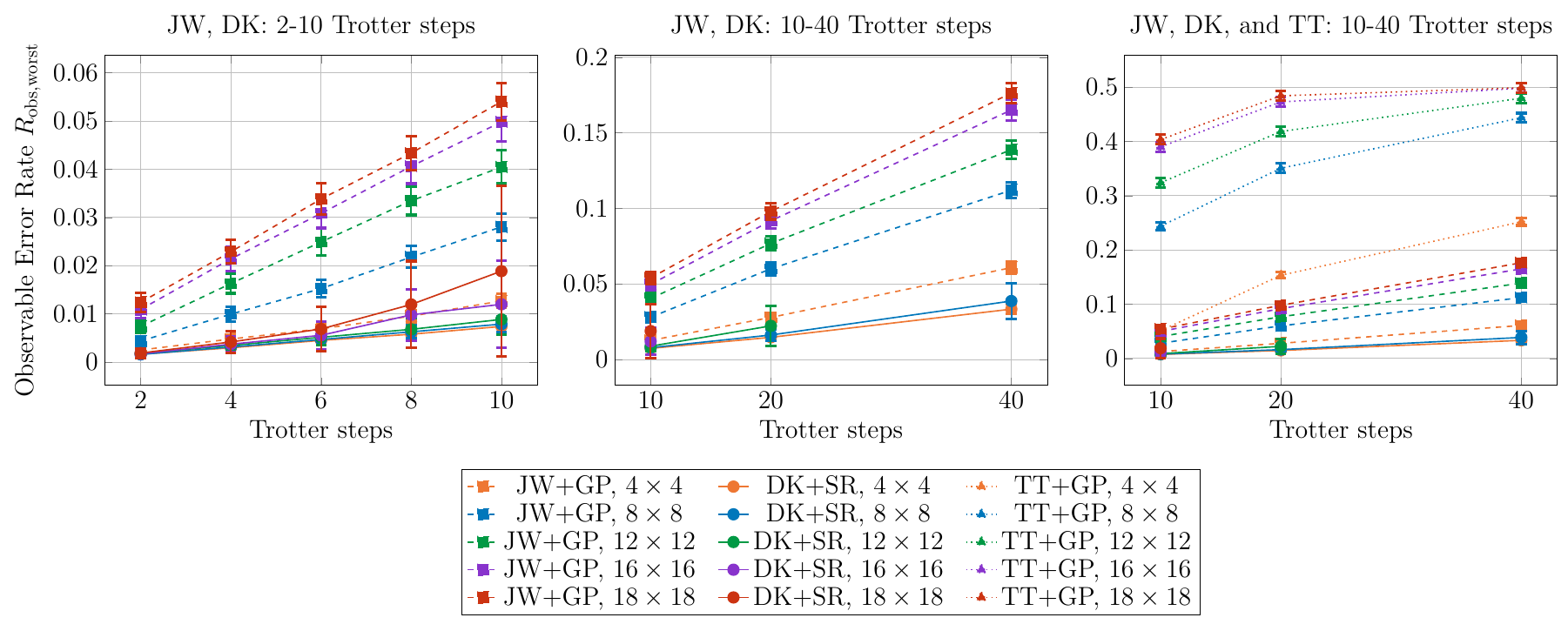}
        \label{fig:enc-comparison-si-b}
        }
    \caption{
    \textbf{Performance of fermionic encodings in the superconducting inspired error model.}
    We consider the DK, JW, and TT encodings, \rev{with stabilizer reconstruction (SR) and global parity postselection (GP).}
    One Trotter step corresponds to an application of all the Hamiltonian terms, and we measure the occupation number at the end of the circuit.
    $R_{\mathrm{obs,worst}}$ is the maximum local observable error rate and data points with error detection rate $R_{\text{det}}\geq0.995$ are excluded.
    Error bars are computed using bootstrap resampling with 1000 resamples.
    (a) compares encoding performance across error rates.
    (b) zooms in on the lowest error rate, $p=0.01\%$: the left plot magnifies DK and JW data from (a), the center compares deeper circuits for DK and JW, and the right includes all three encodings (note the y-axis scale change).
    }
     \label{fig:enc-comparison-si}
\end{figure*}

\begin{figure*}
    \centering
   \subfloat[Observable errors for occupation number measurements at $p=0.05\%$ (superconducting inspired error model)]{
        \centering
        \includegraphics[width=0.8\linewidth]{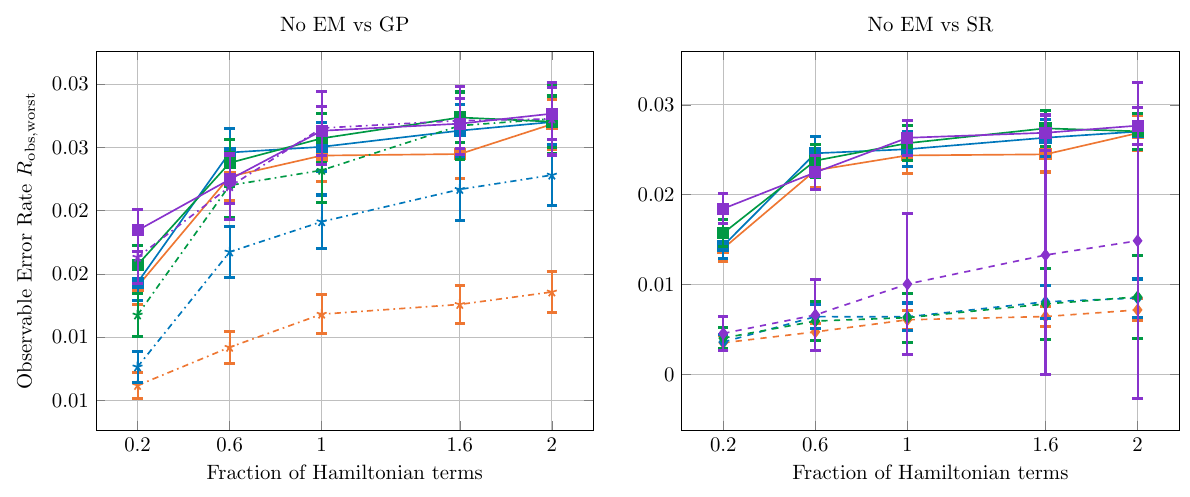}
        \label{fig:compact-obs-occ-si}
    }
    
    \subfloat[Observable errors for hopping term measurements at $p=0.05\%$ (superconducting inspired error model)]{
        \centering
        \includegraphics[width=0.8\linewidth]{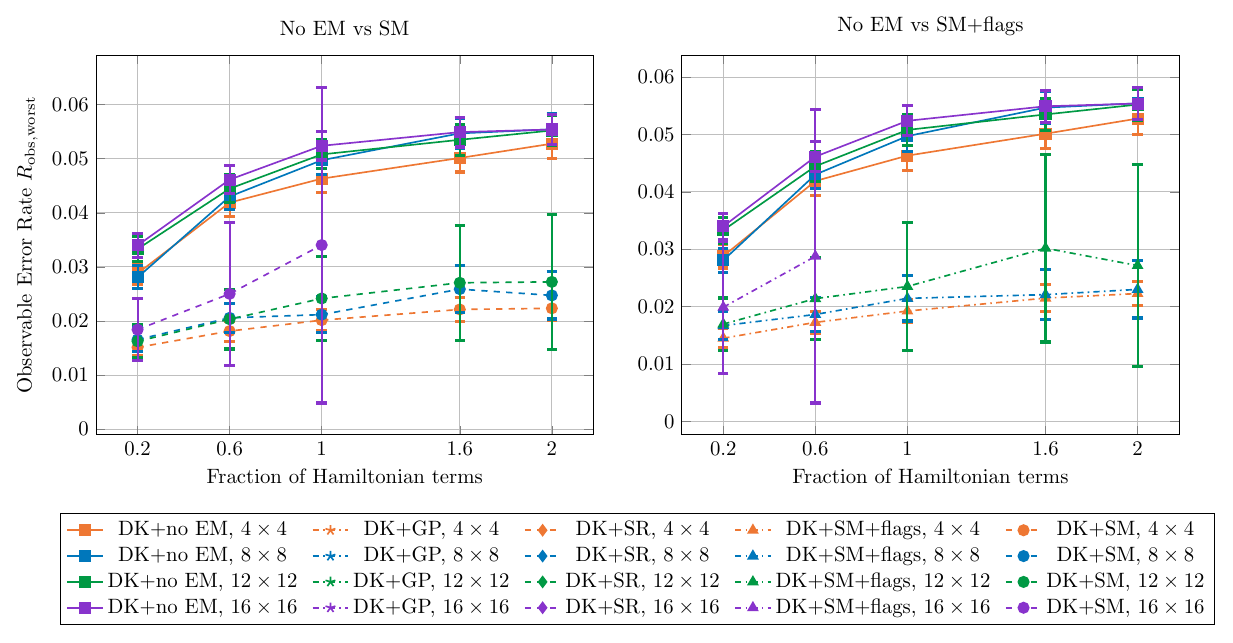}
        \label{fig:compact-obs-hop-si}
    }
    
   \subfloat[Error detection rates at different physical error rates (superconducting inspired error model)]{
        \centering
        \includegraphics[width=0.9\linewidth]{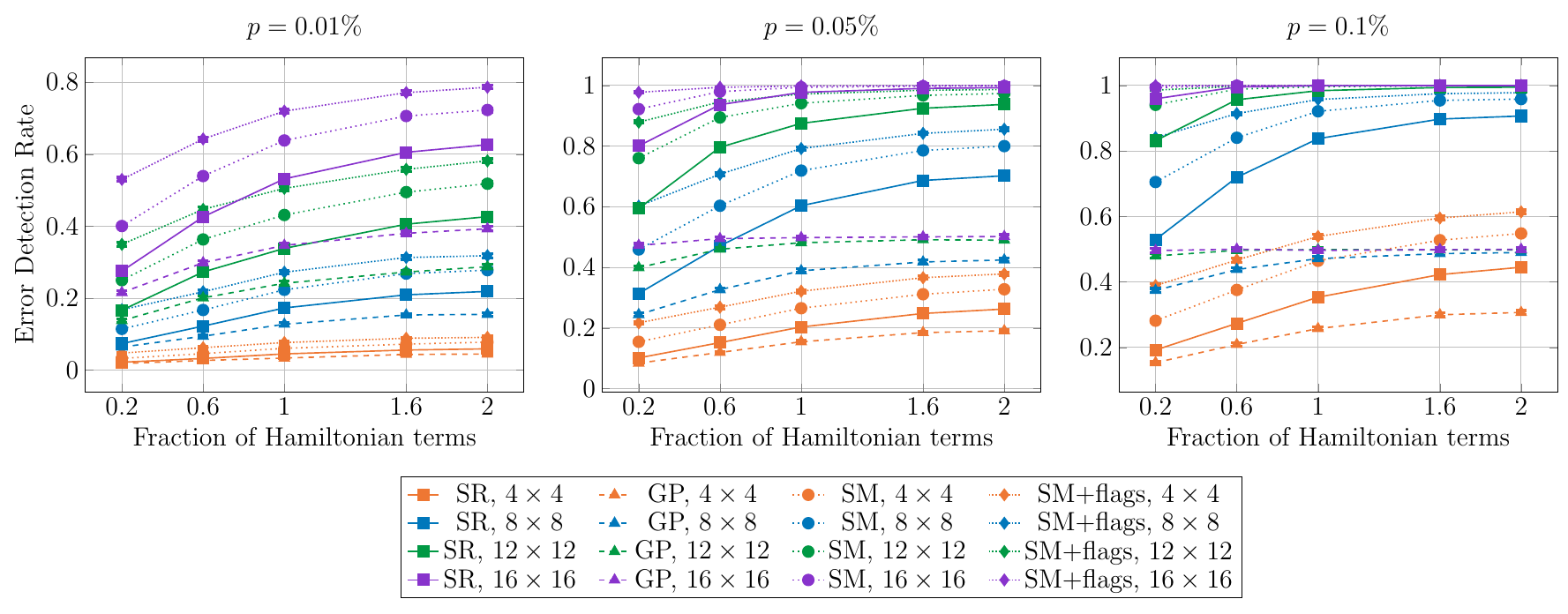}
        \label{fig:compact-det-si}
        }

  \caption{
  \textbf{Performance of various postselection methods in the DK encoding (SI error model).} 
  Each circuit applies a random sequence of logical operators, with operator count given as a fraction of Hamiltonian terms.
  Error bars are computed using bootstrap resampling with 1000 resamples.
  (a) and (b) showcase improvements in accuracy enabled by global parity postselection (GP) and stabilizer reconstruction (SR) for occupation number measurements, as well as stabilizer measurement (SM) and stabilizer measurement with flag qubits (SM+flags) for hopping term measurements, compared with no error mitigation (no EM). 
  Data points with error detection rate $R_{\mathrm{det}} \geq 0.995$ are excluded.
  The legend in (b) is shared by (a) and (b).
  (c) presents a comparison of the error detection rates of different postselection methods for varying error rates.
  }
    \label{fig:compact-comparison-si}
\end{figure*}

\begin{figure*}
    \centering
    \subfloat[Standard depolarizing (SD) error model]{
        \includegraphics[width=0.9\linewidth]{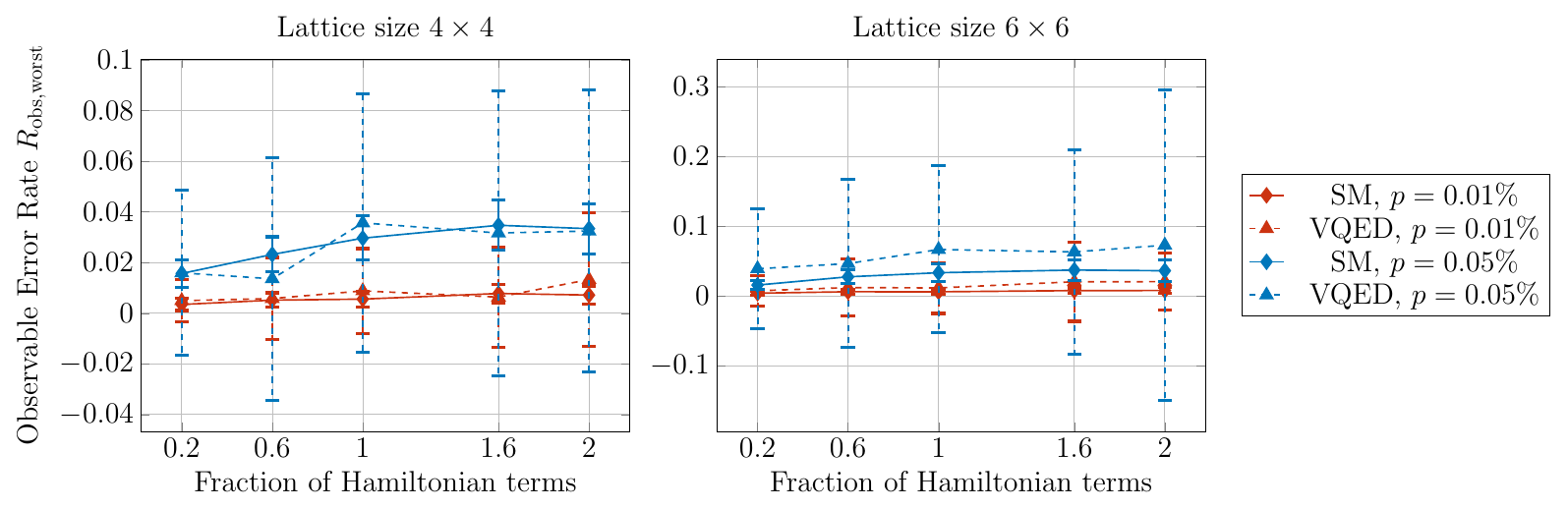}
        \label{fig:vqed_sd6_compare}
    } \\[1em]
    \subfloat[Superconducting inspired (SI) error model]{
        \includegraphics[width=0.9\linewidth]{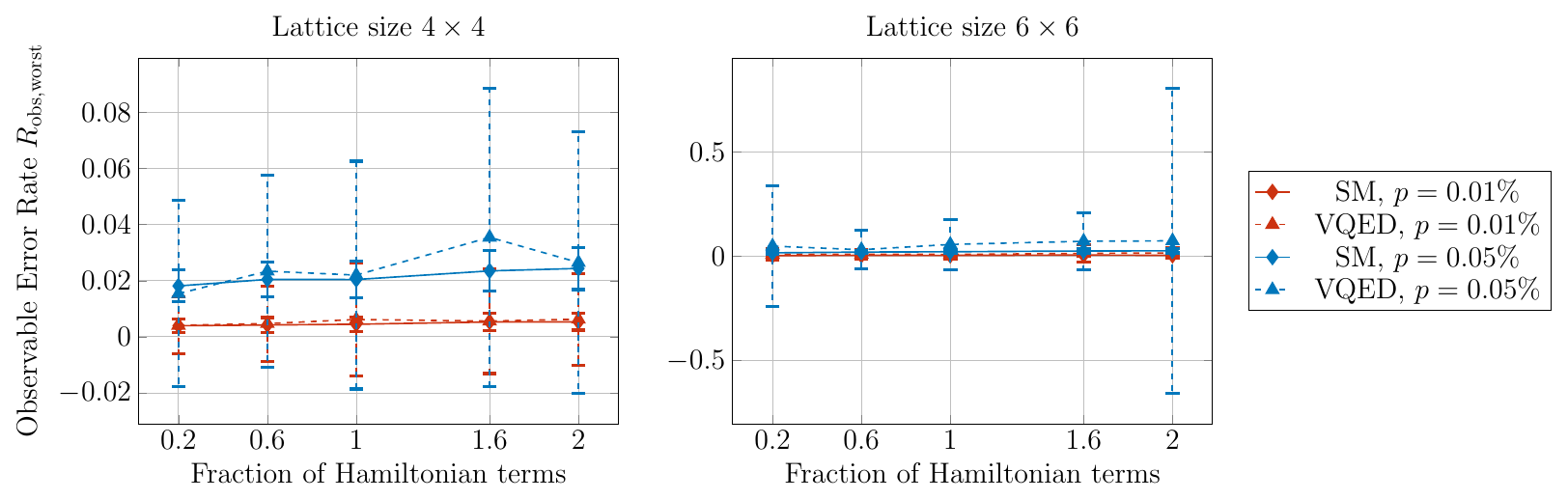}
        \label{fig:vqed_si_compare}
    }
    \caption{
   \rev{\textbf{Comparison of virtual quantum error detection (VQED) and stabilizer measurement (SM) for the DK encoding under two physical error models.}
    Each circuit applies a random sequence of logical operators, with operator count given as a fraction of Hamiltonian terms. Final measurements are performed in the hopping operator basis, and \( R_{\mathrm{obs,worst}} \) denotes the maximum local observable error rate.
    For the SM data, error bars are obtained via bootstrap resampling with 10,000 resamples, while for the VQED data, error bars indicate the variance of the estimate.
    }}
    \label{fig:vqed_combined}
\end{figure*}

\textbf{Superconducting inspired error model.} 
\cref{fig:enc-comparison-si,fig:compact-comparison-si,fig:vqed_si_compare} are the same as the figures as in main text but with the SI error model. 
This model, with its reduced single-qubit error rate, provides a slight advantage for DK encoding with 20 Trotter steps potentially accessible for a $12\times12$ lattice at the lowest physical error rate of $p=0.01\%$ (central plot of~\cref{fig:enc-comparison-si-b}). We see no significant differences in the performance of various error mitigation methods for the DK encoding in~\cref{fig:compact-comparison-si}. The performance of JW and TT encodings are almost identical to the SD error model case.

\begin{figure*}
    \centering
        \subfloat[Standard depolarizing error model at different error rates $p$]{
        \centering
        \includegraphics[width=0.95\linewidth]{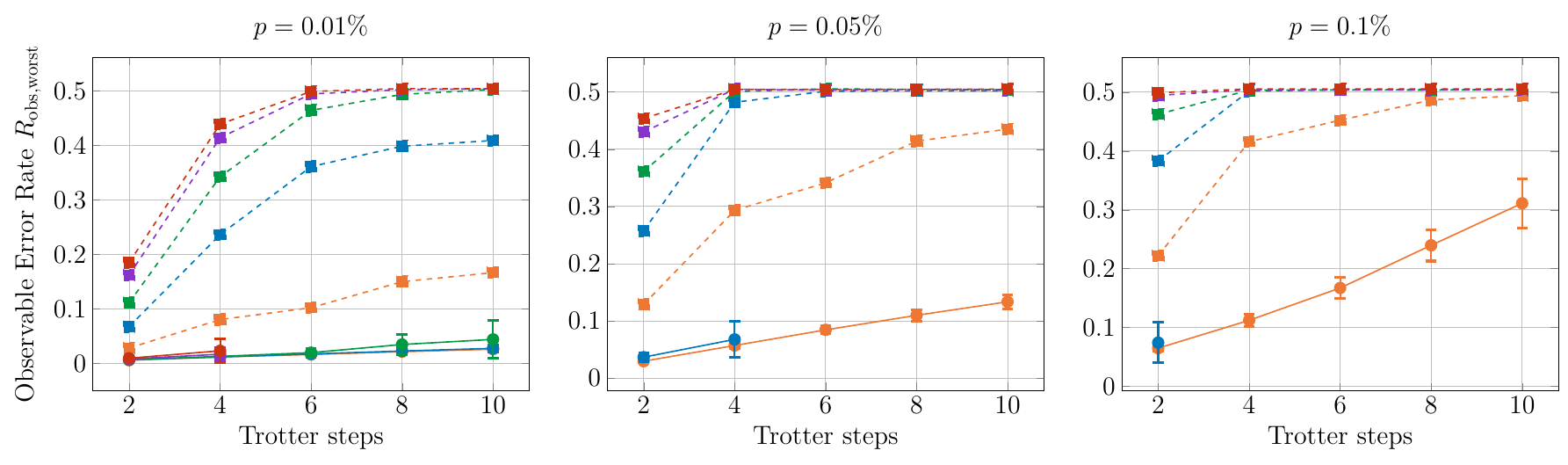}
        
    \label{fig:hop-sd-enc-comparison}
    }

   \subfloat[Superconducting inspired error model at different error rates $p$]{
        \centering
        \includegraphics[width=0.95\linewidth]{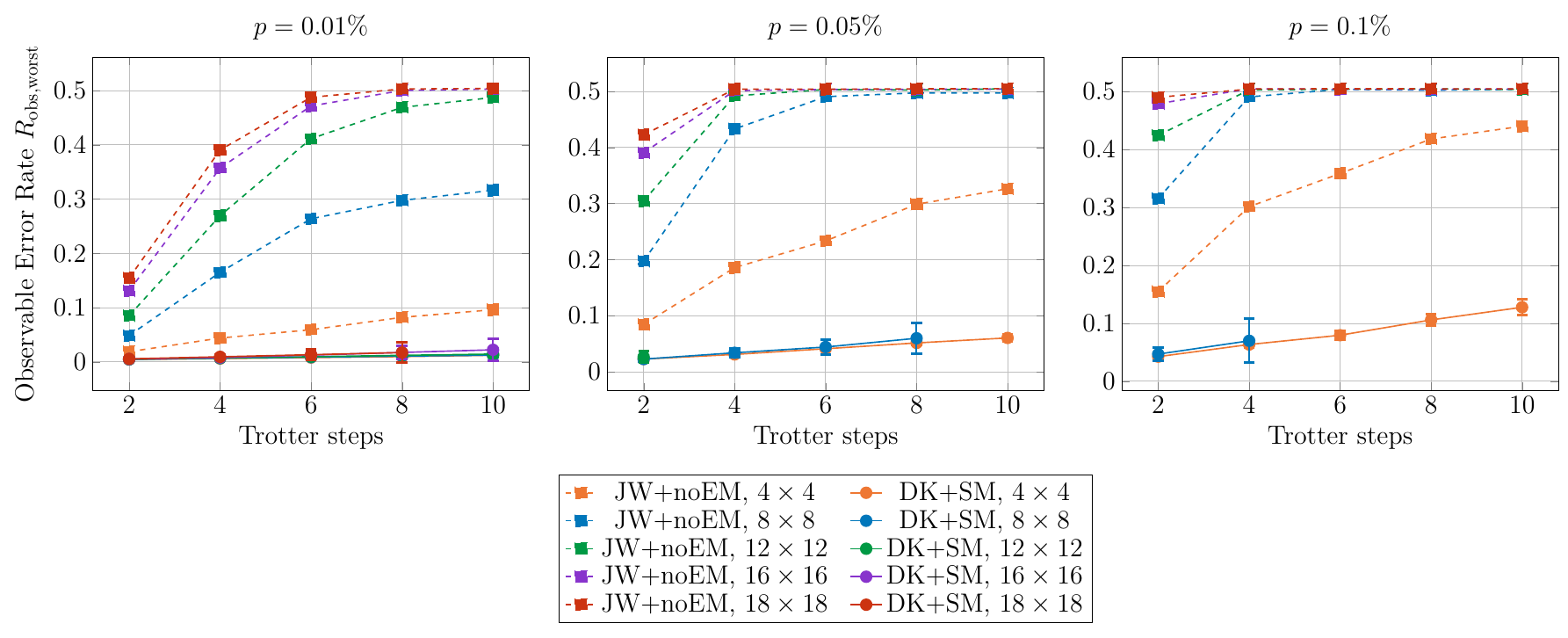}
        \label{fig:hop-si-enc-comparison}
        }
    \caption{\textbf{Comparison of hopping term measurement performance at different error rates.}
    (a) We consider the DK encoding with stabilizer measurement (DK+SM) and the JW encoding (with no error mitigation) on fermionic lattice sizes from $4\times 4$ to $18\times 18$. One Trotter step corresponds to an application of all the terms in the Hamiltonian and the we measure in the hopping operator basis at the end of each circuit.
    $R_{\mathrm{obs,worst}}$ is the maximum local observable error rate and data points with error detection rate $R_{\text{det}}\geq0.995$ are excluded.
    Error bars are computed using bootstrap resampling with 1000 resamples.
    We simulate the standard depolarizing error model with strength $p=0.01\%$ (left), $p=0.05\%$ (middle), and $p=0.1\%$ (right).
    (b) Identical to (a) except with the superconducting inspired error model.
    }
    \label{fig:hop-enc-comparison}
\end{figure*}

\textbf{Hopping term measurements.} 
\cref{fig:hop-enc-comparison} provides a similar comparison of JW and DK to~\cref{fig:enc-comparison,fig:enc-comparison-si} but with measurements extracting hopping term observables. 
For DK circuits we measure all hopping terms colored in red in~\cref{fig:ce-hopping-measurement} with stabilizer measurement for postselection (no flag qubits). 
For JW, we measure $X_iX_{i+1}$ observables, which is the most efficient (in terms of the number of Hamiltonian terms extracted) naive hopping term measurement. 
We do not include results for TT here because we do not have a comparable measurement strategy available for it.
We observe that this kind of measurement significantly disadvantages the JW encoding (with fermionic swap networks) compared with the occupation number measurement.
Hence, the difference in performance between DK and JW is much more pronounced, with the potential to extract relatively low-error observables using the DK encoding that are impossible using the JW encoding.  
However, the DK encoding suffers from high postselection costs in this setting as well, with most promising results being 10 Trotter steps for the $16\times16$ lattice and 8 Trotter steps for the $18\times18$ lattice.

\begin{figure*}
    \centering
        \subfloat[Occupation number measurement]{
        \centering
        \includegraphics[width=0.95\linewidth]{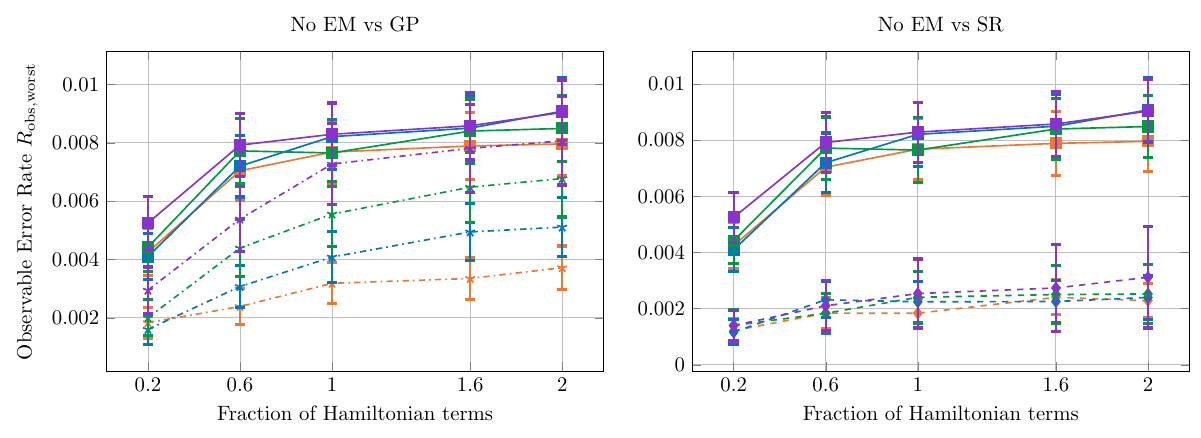}
    \label{fig:dk-sd-err0001-occ}
    }

   \subfloat[Hopping term measurement]{
        \centering
        \includegraphics[width=0.95\linewidth]{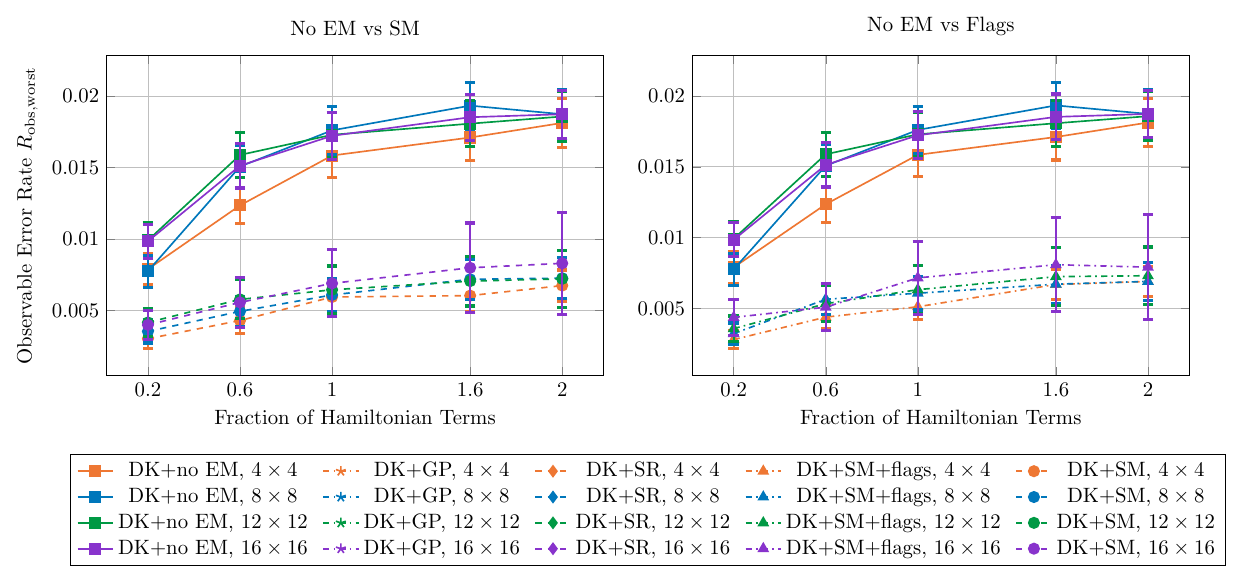}
        \label{fig:dk-sd-err0001-hop}
        }
    \caption{
    \textbf{Performance of the DK encoding in the aspirational error regime ($p=0.01\%$) for the standard depolarizing error model.}
    We consider fermionic lattice sizes of $4\times 4$ up to $16 \times 16$.
    Each circuit applies a random sequence of logical operators, with operator count given as a fraction of Hamiltonian terms.
    $R_{\mathrm{obs,worst}}$ is the maximum local observable error rate and data points with error detection rate $R_{\text{det}}\geq0.995$ are excluded.
    Error bars are computed using bootstrap resampling with 1000 resamples.
    (a) Performancce comparison of global parity postselection (GP) and stabilizer reconstruction (SR) with no error mitigation (no EM) for occupation number measurements.
    (b) Performance comparison of stabilizer measurement (SM) and flag qubits with no error mitigation for hopping term measurements.
    }    
    \label{fig:dk-err0001-sd}
\end{figure*}

\begin{figure*}
    \centering
        \subfloat[Occupation number measurement]{
        \centering
        \includegraphics[width=0.95\linewidth]{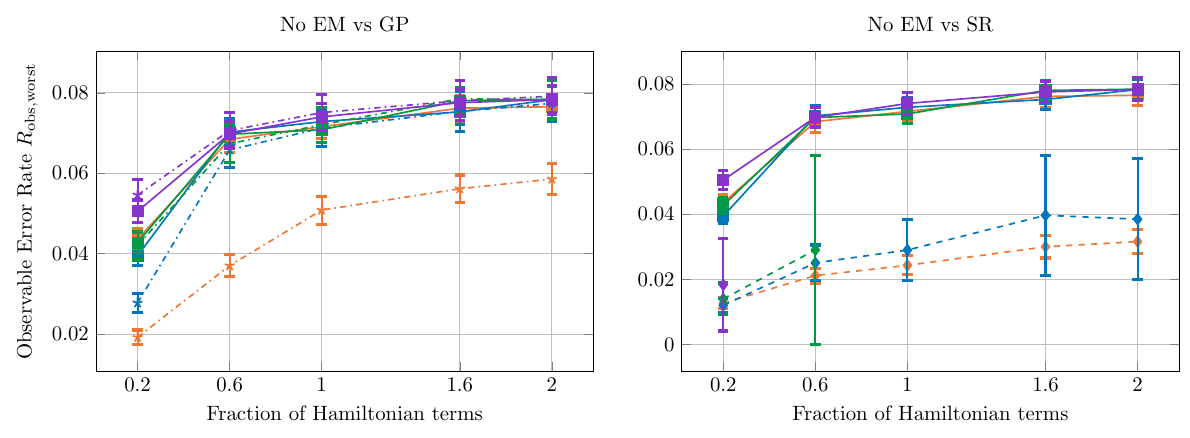}
    \label{fig:dk-sd-err001-occ}
    }

   \subfloat[Hopping term measurement]{
        \centering
        \includegraphics[width=0.95\linewidth]{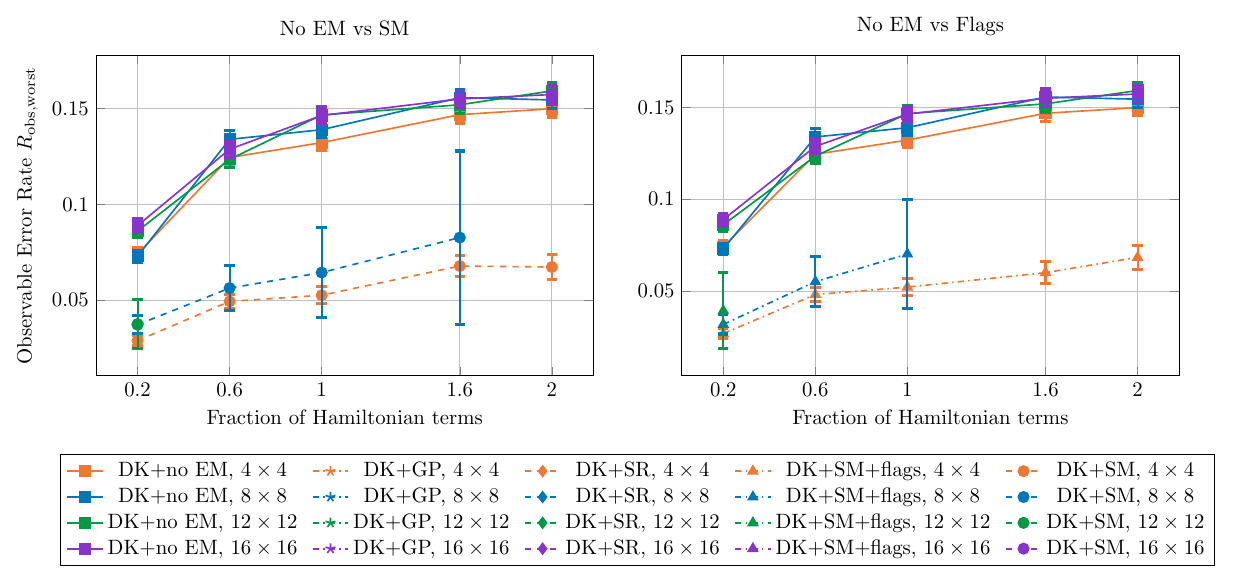}
        \label{fig:dk-sd-err001-hop}
        }
    \caption{
    \textbf{Performance of the DK encoding in the near future error regime ($p=0.1\%$) for the standard depolarizing error model.}
    We consider fermionic lattice sizes of $4\times 4$ up to $16 \times 16$.
    Each circuit applies a random sequence of logical operators, with operator count given as a fraction of Hamiltonian terms.
    $R_{\mathrm{obs,worst}}$ is the maximum local observable error rate and data points with error detection rate $R_{\text{det}}\geq0.995$ are excluded.
    Error bars are computed using bootstrap resampling with 1000 resamples.
    (a) Performancce comparison of global parity postselection (GP) and stabilizer reconstruction (SR) with no error mitigation (no EM) for occupation number measurements.
    (b) Performance comparison of stabilizer measurement (SM) and flag qubits with no error mitigation for hopping term measurements.
    }    
    \label{fig:dk-err001-sd}
\end{figure*}

\begin{figure*}
    \centering
        \subfloat[Occupation number measurement]{
        \centering
        \includegraphics[width=0.95\linewidth]{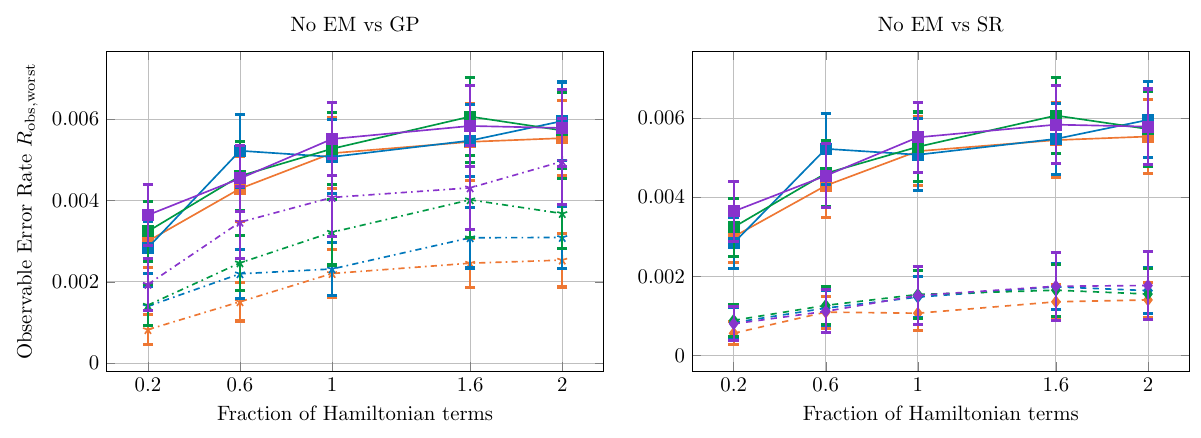}
    \label{fig:dk-si-err0001-occ}
    }

   \subfloat[Hopping term measurement]{
        \centering
        \includegraphics[width=0.95\linewidth]{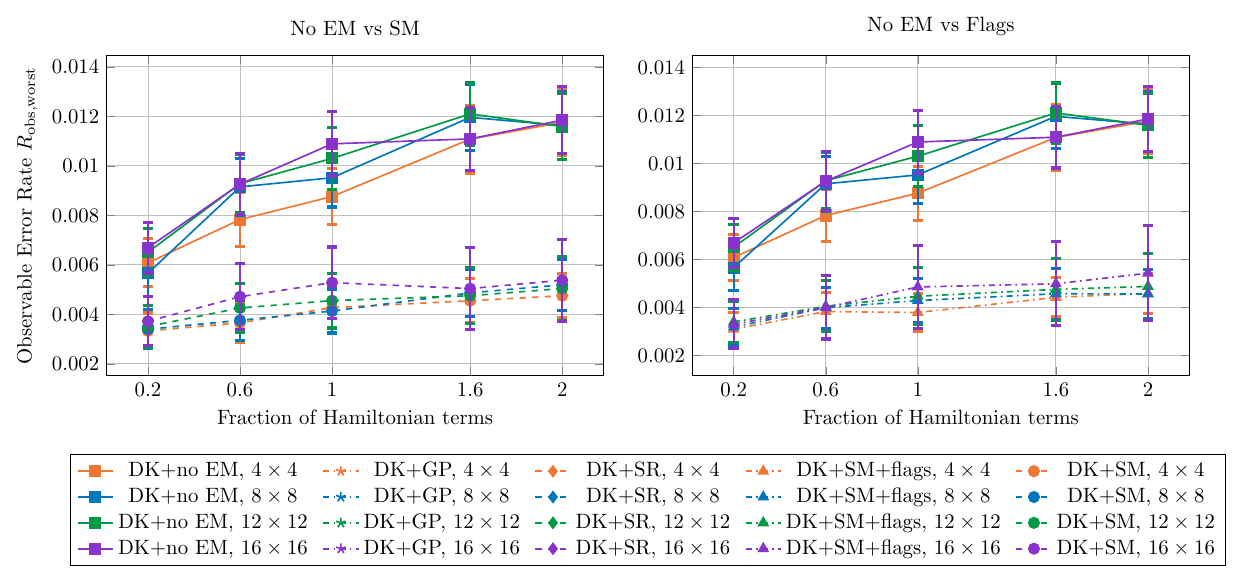}
        \label{fig:dk-si-err0001-hop}
        }
    \caption{
    \textbf{Performance of the DK encoding in the aspirational error regime ($p=0.01\%$) for the superconducting inspired error model.}
    We consider fermionic lattice sizes of $4\times 4$ up to $16 \times 16$.
    Each circuit applies a random sequence of logical operators, with operator count given as a fraction of Hamiltonian terms.
    $R_{\mathrm{obs,worst}}$ is the maximum local observable error rate and data points with error detection rate $R_{\text{det}}\geq0.995$ are excluded.
    Error bars are computed using bootstrap resampling with 1000 resamples.
    (a) Performancce comparison of global parity postselection (GP) and stabilizer reconstruction (SR) with no error mitigation (no EM) for occupation number measurements.
    (b) Performance comparison of stabilizer measurement (SM) and flag qubits with no error mitigation for hopping term measurements.
    }    
    \label{fig:dk-err0001-si}
\end{figure*}

\begin{figure*}
    \centering
        \subfloat[Occupation number measurement]{
        \centering
        \includegraphics[width=0.95\linewidth]{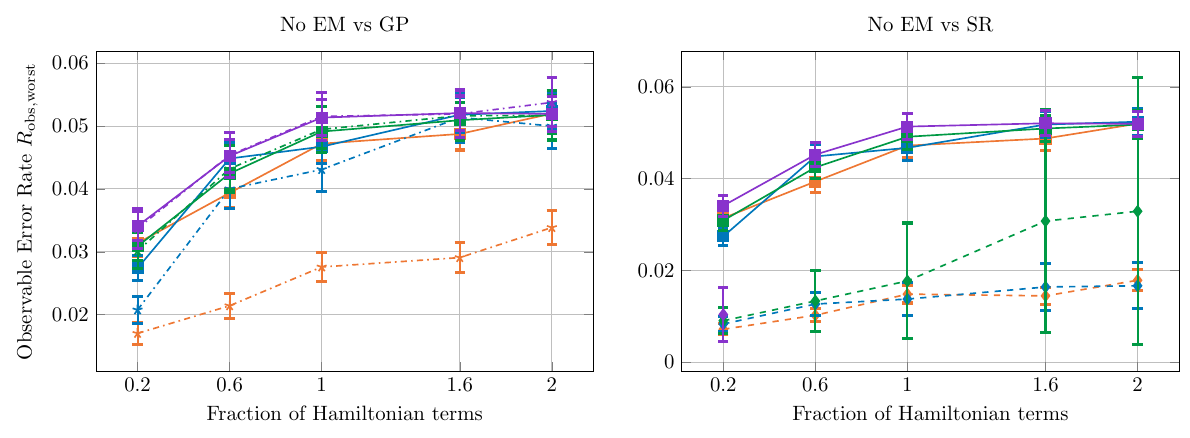}
    \label{fig:dk-si-err001-occ}
    }

   \subfloat[Hopping term measurement]{
        \centering
        \includegraphics[width=0.95\linewidth]{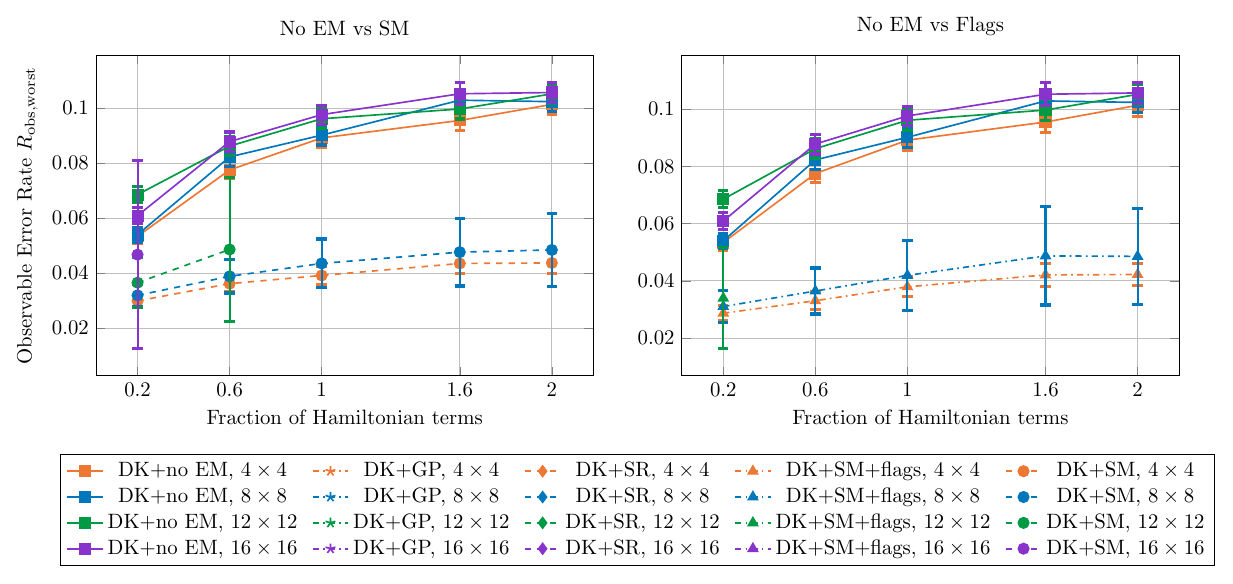}
        \label{fig:dk-si-err001-hop}
        }
    \caption{
    \textbf{Performance of the DK encoding in the near future error regime ($p=0.1\%$) for the superconducting inspired error model.}
    We consider fermionic lattice sizes of $4\times 4$ up to $16 \times 16$.
    Each circuit applies a random sequence of logical operators, with operator count given as a fraction of Hamiltonian terms.
    $R_{\mathrm{obs,worst}}$ is the maximum local observable error rate and data points with error detection rate $R_{\text{det}}\geq0.995$ are excluded.
    Error bars are computed using bootstrap resampling with 1000 resamples.
    (a) Performancce comparison of global parity postselection (GP) and stabilizer reconstruction (SR) with no error mitigation (no EM) for occupation number measurements.
    (b) Performance comparison of stabilizer measurement (SM) and flag qubits with no error mitigation for hopping term measurements.}    
    \label{fig:dk-err001-si}
\end{figure*}

\textbf{DK comparison for more error rates (both error models).}
\cref{fig:compact-comparison,fig:compact-comparison-si} compare the DK encoding in the intermediate error regime of $p=0.05\%$ of SD and SI error models respectively. 
As mentioned in the main text, we highlight this error regime as it best demonstrates the tradeoff between shot counts and accuracy in the DK encoding. 
\cref{fig:dk-err0001-sd,fig:dk-err001-sd,fig:dk-err0001-si,fig:dk-err001-si} demonstrate the performance of the DK encoding in the aspirational ($p=0.01\%$) and near future ($p=0.1\%$) regimes for both the SD and SI error models.
We again see no notable differences between SD and SI error models.
Results in both regimes show little to no advantage from GP compared to no error mitigation, with some improvements present only in smaller lattices and low error rates. 
Meanwhile, when there are enough samples to discard in the aspirational regime, stabilizer reconstruction and stabilizer measurements provide an improvement in accuracy for all lattice sizes. 
However, we see that the sampling requirements are impractical for the near future error regime. 

\begin{figure*}
    \centering
    \includegraphics[width=1\linewidth]{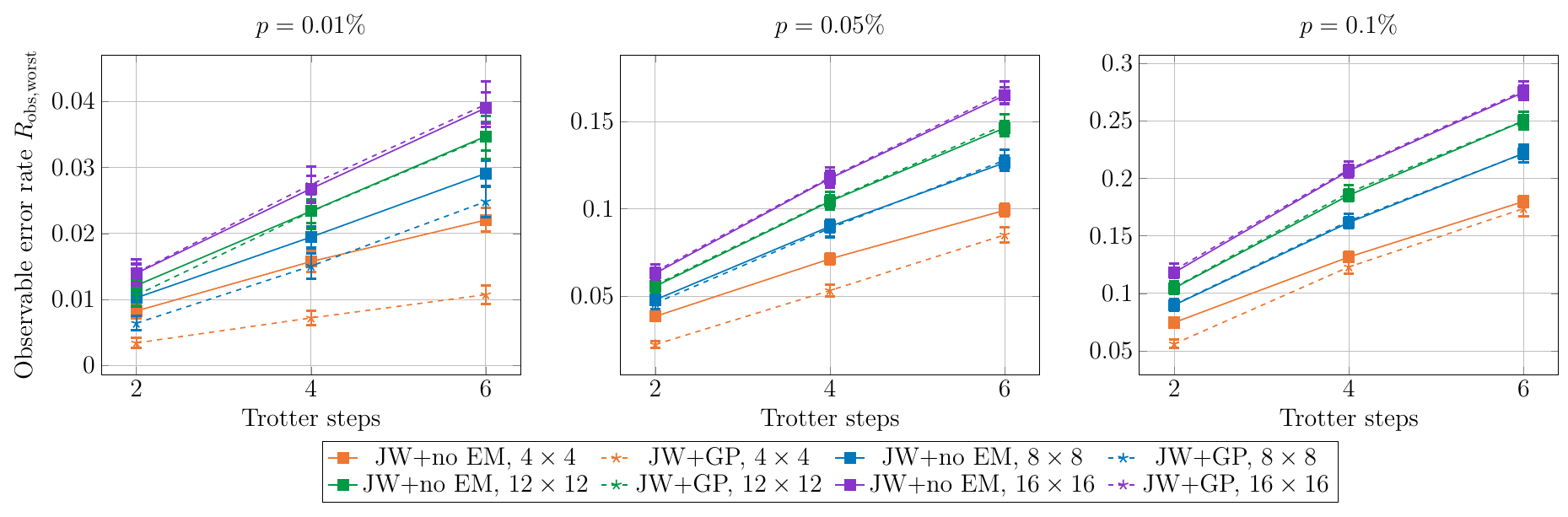}
    \caption{
    \textbf{Effect of global parity postselection (GP) in the JW encoding for the standard depolarizing error model.}
    We consider fermionic lattice sizes of $4\times 4$ up to $16 \times 16$ and occupation number measurements.
    One Trotter step consists of an application of all terms in the Hamiltonian and $R_{\mathrm{obs,worst}}$ is the maximum local observable error rate.
    Error bars are computed using bootstrap resampling with 1000 resamples.
    }
    \label{fig:jw-gp-compare}
\end{figure*}

\textbf{JW with GP vs. no GP.} \Cref{fig:jw-gp-compare} demonstrates no significant improvement in accuracy resulting from GP in JW circuits except for the smallest lattice sizes and lowest error rates. 

\rev{\section{Virtual quantum error detection}}
\label{app:vqed}
\rev{In a recent work~\cite{tsubouchi2023virtual}, it was shown that error detection can be performed by measuring a constant number of stabilizer generators rather than a generating set.
This technique is termed virtual quantum quantum error detection (VQED) and is a generalization of symmetry expansion~\cite{mcclean2020decoding,cai2021quantum}.
VQED is defined in terms of an observable, $\mathcal O$ measured at the end of the simulation.}

\rev{To perform VQED, we replace measurement of the stabilizer generators with the circuit shown in \cref{fig:vqed}, where the stabilizers $S_j$ and $S_k$ are chosen randomly.
Let $m$ be the number of VQED circuits in the overall circuit and suppose we perform $N$ total runs.
For run $j$, let $b_j$ be the product of the measurement outcomes of the auxiliary qubits for each of the $m$ rounds of VQED and let $o_j$ be the product of the measurement outcome of $\mathcal O$ with $b_j$.
Our estimate of the expectation value of $\mathcal O$ is then
\begin{equation}
    \widetilde{\langle \mathcal{O}\rangle} = \left. \sum_{j=1}^N o_j \middle/ \sum_{j=1}^N b_j \right.
    .
\end{equation}
The overhead in this technique is inverse quadratic in the target error $\epsilon$ and the trace of state prior to measurement~\cite{tsubouchi2023virtual}.}

\begin{figure}[ht]
    \centering
    \begin{tikzpicture}
        \node[scale=1] {
        \begin{quantikz}
            \lstick{$|\psi\rangle$} & \gate{S_j} & \gate{S_k} & \\
            \lstick{$|+\rangle$} && \ctrl{-1} & \gate{H} & \meter{} \\
        \end{quantikz}
        };
    \end{tikzpicture}
    \caption{\rev{Virtual quantum error detection circuit. $|\psi\rangle$ is an encoded state and $S_j$ are stabilizer generators.}}
    \label{fig:vqed}
\end{figure}

\rev{For the DK encoding, we can replace the control-$S_k$ in \cref{fig:vqed} with the flagged version in \cref{fig:stab-circuit} to ensure robustness.
We also note that VQED assumes an input state that is the $+1$ eigenstate of the stabilizer generators. 
If instead that the input state has $-1$ eigenvalue for stabilizer $S_j$, we can still use the circuit in \cref{fig:vqed} but now we must multiply our measurement outcome by $-1$ before including it in the $o_j$ product.}

\rev{We assess the feasibility of VQED for the DK encoding in~\cref{fig:vqed_combined} for random logical circuits and hopping term measurements. Our results suggest that the sampling overhead scaling of VQED makes it unsuitable for this application, even for relatively small lattice sizes. This is reflected in the magnitude of error bars, caused by the high variance of results.
Both VQED and error mitigation via syndrome extraction circuits were tested in a reduced sample regime of $n_{\mathrm{samp}} = 10{,}000$ due to the high classical simulation cost of VQED, which results from randomized circuit generation. We observe that for the same number of samples, traditional syndrome extraction yields lower error in observable readouts with reduced variance.}

\end{document}